\documentclass[11pt]{article}

\usepackage{epsfig}
\usepackage{amssymb}
\usepackage{times}
\usepackage{amsmath}
\usepackage{comment}

\title{Existentially Restricted Quantified Constraint Satisfaction}
\author{Hubie Chen\\Universitat Pompeu Fabra\\Barcelona, Spain\\{\tt hubie.chen@upf.edu}}
\date{ }

\newtheorem{theorem}                            {Theorem}[section]
\newtheorem{lemma}              [theorem]       {Lemma}

\newtheorem{corollary}          [theorem]       {Corollary}
\newtheorem{definition}         [theorem]       {Definition}
\newtheorem{prop}               [theorem]       {Proposition}
\newtheorem{obs}               [theorem]        {Observation}
\newtheorem{example}		[theorem]	{Example}

\newenvironment{pf}{\noindent\textbf{Proof\/}.}{$\Box$ \vspace{1mm}}

\newcommand{\qcsp}{\ensuremath{\mathsf{QCSP}}}
\newcommand{\tqcsp}{\ensuremath{\qcsp_t}}
\newcommand{\csp}{\ensuremath{\mathsf{CSP}}}

\newcommand{\cons}{\ensuremath{\mathcal{C}}}

\newcommand{\qcspe}{\ensuremath{\qcsp^{\exists}}}
\newcommand{\tqcspe}{\ensuremath{\qcspe_t}}

\newcommand{\pol}{\mathsf{Pol}}
\newcommand{\inv}{\mathsf{Inv}}

\newcommand{\alga}{\mathcal{A}}
\newcommand{\algb}{\mathcal{B}}

\newcommand{\rela}{\mathbf{A}}
\newcommand{\relb}{\mathbf{B}}
\newcommand{\relc}{\mathbf{C}}
\newcommand{\relbg}{\relb^{\Gamma}}

\newcommand{\fcoll}{\mathcal{F}}

\newcommand{\sigmap}{\Sigma^{\mathsf{p}}}
\newcommand{\pip}{\Pi^{\mathsf{p}}}
\newcommand{\conp}{\ensuremath{\mathsf{coNP}}}
\newcommand{\np}{\ensuremath{\mathsf{NP}}}
\newcommand{\pspace}{\ensuremath{\mathsf{PSPACE}}}
\newcommand{\p}{\ensuremath{\mathsf{P}}}

\newcommand{\lexpr}{\ensuremath{\langle}}
\newcommand{\rexpr}{\ensuremath{\rangle}}

\newcommand{\od}{\mathcal{O}_D}
\newcommand{\rd}{\mathcal{R}_D}

\newcommand{\pow}{\wp}
\newcommand{\fcollp}{\fcoll_{\pow(D)}}

\begin{document}

\maketitle

\begin{abstract}
The quantified constraint satisfaction problem (QCSP) is a powerful
framework for modelling computational problems.
The general intractability of the QCSP has motivated the pursuit
of restricted cases that avoid its maximal complexity.
In this paper, we introduce and study a new model for investigating
QCSP complexity
in which the types of constraints given by the existentially
quantified variables, is restricted.
Our primary technical contribution is the development and application
of a general technology for proving positive results
on parameterizations of the model,
of inclusion
in the complexity class coNP.
\end{abstract}

\section{Introduction}

\subsection{Background}

The \emph{constraint satisfaction problem (CSP)} is a general framework
in which many combinatorial search problems can be conveniently formulated.
Intuitively, the CSP involves deciding if a collection of constraints
on a set of variables can be simultaneously satisfied.
The CSP can be formalized as the problem of deciding the truth of a
first-order sentence
consisting of a conjunction of constraints, in front of which all variables
are existentially quantified.

A natural and useful generalization of the CSP is the
\emph{quantified constraint satisfaction problem (QCSP)}.  The definition
of the QCSP is similar to that of the CSP, but variables may be both
universally and existentially quantified.  
While the CSP lies in the complexity class NP and hence can only
be used to model other problems in NP, the higher expressivity
of the QCSP permits the modelling of problems in the (presumably)
larger complexity class PSPACE.  Such problems arise naturally
in a wide variety of domains, for example, logic, artificial intelligence,
verification,
combinatorics, and game theory.

In their general formulation, the CSP and QCSP are intractable,
being NP-complete and PSPACE-complete, respectively; this 
intractability
motivates the pursuit of restricted cases of these problems that
avoid ``maximal'' complexity, and
fall into complexity classes strictly below NP and PSPACE, respectively.
It is possible to parameterize these problems
by restricting the \emph{constraint language},
or the types of constraints that are permitted in problem instances.
This form of restriction captures and places into a unified framework
many particular cases of the CSP and QCSP 
that have been independently
investigated, 
including the {\sc Horn Satisfiability} and {\sc 2-Satisfiability} problems,
and their quantified versions.
The notion of constraint language has
 its roots in 
the classic dichotomy theorem of Schaefer
\cite{schaefer78}, which shows that every constraint
language over a two-element domain gives rise to a case of the CSP 
that is either in P, or is NP-complete.
The research program
of classifying the CSP complexity of all constraint languages
over domains of arbitrary finite size has attracted significant attention; see
for instance~\cite{jcg97,fv98,dp99,bkj00,bulatov02-dichotomy,bulatov02-twosemilattices,bulatov02-maltsev,bulatov03,dalmau04-maltsev}.

On the QCSP front, previous work on constraint language restrictions
is as follows.
An analog of Schaefer's theorem 
is known~\cite{dalmau97,cks01},
which shows that all constraint languages over a two-element domain
give rise to a case of the QCSP that is either in P, or is PSPACE-complete.
A ``finer'' version studying the alternation-bounded QCSP over
a two-element domain has also been obtained~\cite{hemaspaandra04}.
Recently, the study of 
constraint languages 
in domains of size larger than two has been initiated~\cite{bbkj03,chen05-thesis,chen05-maximal}.
Results include the development of an algebraic theory for studying
the QCSP~\cite{bbkj03}, general technology for proving positive complexity
results~\cite{chen05-thesis}, and some broad
 classification results~\cite{chen05-maximal}.

\subsection{A New Model: Existentially Restricted Quantified Constraint Satisfaction}

In all previous work on QCSP complexity,
constraint language restrictions are applied
equally 
to both universally and existentially quantified variables; 
that is, in constraints,
for any position where an existentially quantified variable can occur,
a universally quantified variable can also occur in that position, and
vice-versa.
This paper introduces and studies a \emph{new model} for investigating
QCSP complexity, where constraint language restrictions are applied
\emph{only} to existentially quantified variables.
We call our new model
\emph{existentially restricted quantified constraint satisfaction},
and
 refer to the previously studied QCSP 
model simply as the \emph{standard model}.

Both our new model and the standard model are generalizations of the
usual CSP model: when these two models are
restricted to instances
having only existential quantification, they coincide and yield the
CSP model.  
However, there is a principal difference between these two QCSP models
which rears its head as soon as universal quantification is permitted:
although it is possible to obtain polynomial-time
tractability results in the standard model for interesting constraint
languages,
under extremely mild assumptions on the constraint
language, our new model is at least coNP-hard.
This, of course, reflects the definition of our model, in which the
universal variables do not need to observe any form of restriction.
In consequence, the best type of positive complexity result
that one can reasonably hope for in our new model is a demonstration of
containment inside coNP.  Accordingly, the main technical contribution
of this paper is the development and application of a general
technology for proving coNP-inclusion results in our new model.

Our new model, as with the standard model and the CSP model,
constitutes a \emph{simple, syntactic} means of restricting a
generally intractable computational problem.
Although this paper is the first to systematically investigate this model,
we view it as being at least as natural as the standard model.
We believe that this model gives rise to beautiful theory at the 
interface of logic, algebra, and computational complexity.
We now turn to articulate some concrete reasons for
interest in our new model.

First, our model allows us to obtain positive complexity results--of 
inclusion in coNP--for classes of QCSP instances for which 
the only complexity result
that can be derived in the standard model is
 the trivial PSPACE upper bound.  Roughly speaking,
this is because there are QCSP instances where the
 constraint language of the existential
variables has tractable structure, but the overall
 constraint language lacks tractable structure.
One example is
the class of \emph{extended quantified Horn formulas}, 
defined by Kleine B\"{u}ning et al.~\cite{bkf95}
as the boolean (two-element) QCSP 
where all constraints must be clauses that,
when restricted to existential
variables, are Horn clauses.
A second example is the boolean QCSP where each
constraint must be a clause in which there are at most two
existential variables; this class forms a natural generalization
of the classical {\sc 2-Satisfiability} problem.
We in fact obtain the first non-trivial complexity upper bounds 
for both of these classes in this paper.

In addition, there are situations in which our new model
can be used to derive \emph{exact} complexity analyses for
constraint languages under the standard model.
Previous work has revealed that there are constraint languages that are 
coNP-hard under the standard model~\cite{chen05-maximal}.
We obtain coNP-inclusion results, and hence coNP-completeness results,
for some of these constraint languages 
\emph{in the standard model, via our new model.}
In particular, we obtain--in the standard model--the 
first non-trivial upper bounds for
constraint languages having a \emph{set function polymorphism},
a robust class of constraint languages that,
in a precise sense, capture the \emph{arc consistency} algorithm
that has been studied heavily in constraint satisfaction~\cite{dp99}.
These standard model coNP-inclusion results are obtained by
first performing a reduction to our new model, and then by establishing
a coNP inclusion in our new model.

\subsection{Overview of results}  
As we have mentioned,
the primary technical contribution
of this paper is the development and application of a general
technology for proving coNP-inclusion results in our new model.
This technology is centered around a new notion 
which we call \emph{fingerprint}
(presented in Section~\ref{sect:fingerprints}).
Intuitively, a fingerprint is a succinct representation of a 
conjunction of constraints.
We require fingerprints to have a number of properties, and highlight
two of these now.
First, we require that
 there is a polynomial-time algorithm that, given
a conjunction of constraints,
computes a fingerprint that represents the constraints.
Second, fingerprints must
 ``encode sufficient information'': from a fingerprint
representing a conjunction of constraints, 
it must be possible to construct
a satisfying assignment for the conjunction
 (assuming that the conjunction was satisfiable in the
first place).
We will say that a 
constraint language for which there is a
set of fingerprints obeying these conditions as well
as some further conditions has a \emph{fingerprint scheme}.

Our goal of proving coNP-inclusion results leads naturally to the 
idea of a proof system in which proofs certify the falsity of
QCSP instances.
After introducing the concept of a fingerprint scheme, 
we indeed present a proof system
applicable to 
any QCSP instance over a constraint language having 
a fingerprint scheme (Section~\ref{sect:proof-system}).
The key feature of this proof system is that it supports
polynomially succinct proofs.  Using this proof system, we show that 
for any constraint language having a fingerprint scheme,
any class of
QCSP instances over the constraint language 
having bounded alternation
is contained in coNP.  
(By bounded alternation, we mean that there is a constant 
that upper bounds
the number of quantifier alternations.)

We then apply the developed technology by showing that a number of 
classes of constraint languages
 have fingerprint schemes, and hence that the described coNP-inclusion
result applies to them (Section~\ref{sect:applications}).
Recent work on 
the complexity of 
constraint satisfaction has heavily exploited the fact that
each constraint language gives rise to a set of 
operations called
\emph{polymorphisms} which
are strongly tied to and can be used to study complexity by means
of algebraic methods~\cite{jcg97,jeavons98,bkj00}.
Correspondingly, 
these classes are described using polymorphisms, and are constituted
of
the constraint languages having
 the following types of polymorphisms: 
set functions, near-unanimity operations, and Mal'tsev operations.
Moreover, for our new model,
we observe a dichotomy theorem for two-element constraint languages: 
they are either in coNP 
under bounded alternation, or of the highest complexity possible
for their quantifier prefix.

Lastly, we study the set function polymorphisms giving rise to
constraint languages that are coNP-hard in the standard model
(Section~\ref{sect:set-functions}).
We use the developed theory to observe that such constraint languages
are in coNP, in the standard model, under bounded alternation.
We then investigate the case of unbounded alternation, and show that
in this case, such constraint languages are $\pip_2$-hard, and thus
\emph{not} in coNP, unless the polynomial hierarchy collapses.
We accomplish this hardness result by showing the $\pip_2$-hardness of
extended quantified Horn formulas, and then reducing from these formulas
to the described constraint languages.
Our $\pip_2$-hardness result on extended quantified Horn formulas
gives the first non-trivial complexity lower bound proved on such formulas.
Finally, we observe 
that--since extended quantified Horn formulas can be captured by our model--the
 $\pip_2$-hardness of these formulas implies that the bounded-alternation
coNP-inclusion results we have obtained cannot, in general, be extended to 
the case of unbounded alternation.

One might summarize the technical contributions of this paper as follows.
We make significant advances in understanding our new model 
in the case of bounded alternation, for which we prove a number
of coNP-inclusion results; and, we establish that the case of
unbounded alternation behaves in a provably different manner.

\section{Preliminaries}

\subsection{Definitions}

Throughout this paper, we use $D$ to denote a \emph{domain}, 
which is a nonempty set of finite size.

\begin{definition}
A \emph{relation} (over $D$) is a subset of $D^k$ for some $k \geq 1$,
and is said to have arity $k$.
A \emph{constant relation} is an arity one relation of size one.
A \emph{constraint} is an expression of the form
$R(w_1, \ldots, w_k)$, where $R$ is an arity $k$ relation 
and the $w_i$ are variables.
A \emph{constraint language} is a set of relations, all of which
are over the same domain.
\end{definition}

An arity $k$ constraint
$R(w_1, \ldots, w_k)$ is true or 
\emph{satisfied} under an interpretation
$f$ defined on the variables $\{ w_1, \ldots, w_k \}$ if
$(f(w_1), \ldots, f(w_k)) \in R$.

\begin{definition}
A \emph{quantified formula} is an expression of the form
$\exists X_1 \forall Y_1 \exists X_2 \forall Y_2 \ldots \exists X_t \cons$
such that 
$t \geq 1$, the sets 
$X_1, Y_1, X_2, \ldots$ are pairwise disjoint sets of variables
called \emph{quantifier blocks}, 
and
none of the sets $X_1, Y_1, X_2, \ldots$ are empty except possibly $X_1$.
Each $X_i$ is called an existential block, and each $Y_i$ is called
a universal block.
The expression $\cons$ is a 
quantifier-free first-order formula with free variables
$X_1 \cup Y_1 \cup X_2 \cup \ldots$.
\end{definition}

We will denote the first existential block of a quantified formula $\phi$
by $X_1^{\phi}$, the first universal block of $\phi$
by $Y_1^{\phi}$, and so forth.
Note that, in this paper, we will primarily consider quantified formulas 
that do not have any free variables.
Truth of a quantified formula is defined as in first-order logic:
a quantified formula is true if there exists an assignment to $X_1$
such that for all assignments to $Y_1$, there exists
an assignment to $X_2$, $\ldots$ such that $\cons$ is true.
A \emph{strategy} for a quantified formula $\phi$ is a sequence
of mappings $\{ \sigma_x \}$ where there is a mapping $\sigma_x$
for each existentially quantified variable $x$ of $\phi$, 
whose range is $D$, and
whose domain is the set of functions mapping from the 
universally quantified variables $Y_x$
preceding $x$, to $D$.
We say that a strategy $\{ \sigma_x \}$
for a quantified formula $\phi$ is a 
\emph{winning strategy} if for all assignments $\tau$ to the 
universally quantified variables of $\phi$ to $D$, when 
each universally quantified variable is set according to $\tau$
and each existentially quantified variable $x$ is set according
to $\sigma_x(\tau|_{Y_x})$, the $\cons$ part of $\phi$ is satisfied.
It is well-known that a quantified formula has a winning strategy
if and only if it is true.

We now define the ``standard model'' of quantified constraint satisfaction.
In this paper, 
the symbol $\Gamma$ will always denote a constraint language.

\begin{definition}
The $\qcsp(\Gamma)$ problem is to decide the truth of a quantified formula
$\exists X_1 \forall Y_1 \exists X_2 \forall Y_2 \ldots \exists X_t \cons$
where $\cons$ is a conjunction of constraints, each of which
has relation from $\Gamma$ and variables from 
$X_1 \cup Y_1 \cup X_2 \cup \ldots$.
\end{definition}

\begin{definition}
For $t \geq 1$, 
the $\tqcsp(\Gamma)$ problem is the restriction of the
$\qcsp(\Gamma)$ problem to instances having $t$ or fewer 
non-empty quantifier blocks.
\end{definition}

\begin{obs}
For all constraint languages $\Gamma$ and $t \geq 1$,
the problem $\tqcsp(\Gamma)$
is 
in the complexity class $\sigmap_{t}$ if $t$ is odd, and
in the complexity class $\pip_{t}$ if $t$ is even.
\end{obs}


The usual CSP model--the class of problems $\csp(\Gamma)$--can 
now be easily defined.

\begin{definition}
The $\csp(\Gamma)$ problem is defined to be
$\qcsp_1(\Gamma)$.
\end{definition}


We now formalize our new model of existentially restricted
quantified constraint satisfaction.  In the standard model,
quantified formulas contain a conjunction of constraints; 
in this new model, quantified formulas contain a conjunction
of \emph{extended constraints}.

\begin{definition}
An \emph{extended constraint} (over $D$) is an expression of the form
$(y_1 = d_1) \wedge \ldots \wedge (y_m = d_m) \Rightarrow R(x_1,\ldots,x_k)$
where $m \geq 0$,
each $y_i$ is a universally quantified variable, 
each $x_i$ is an existentially quantified variable,
each $d_i$ is an element of $D$, and
and $R \subseteq D^k$ is a relation.
\end{definition}

We apply the usual semantics to extended constraints, that is,
an extended constraint is true if $R(x_1,\ldots,x_k)$ is true
or there exists an $i$ such that $y_i \neq d_i$.
Note that we permit $m = 0$, in which case an extended constraint is
just a normal constraint having existentially quantified variables.

Our new model concerns quantified formulas with extended constraints,
that is, formulas of the form
$\exists X_1 \forall Y_1 \exists X_2 \forall Y_2 \ldots \exists X_t \cons$
where $\cons$ is the conjunction of extended constraints.
However, 
the QCSP is typically defined as the problem of deciding such a formula
where $\cons$ is the conjunction of constraints.
We would like to point out that any instance of the QCSP can be converted
to a quantified formula with extended constraints.
Let $R(w_1, \ldots, w_k)$ be a constraint
within a quantified formula, and assume for the sake of notation
that $w_1, \ldots, w_j$ are universally quantified variables and that
$w_{j+1}, \ldots, w_k$ are existentially quantified variables. 
The constraint $R(w_1, \ldots, w_k)$ is semantically equivalent to
the conjunction of the extended constraints
$$(w_1 = d_1) \wedge \ldots \wedge (w_j = d_j) \Rightarrow 
R_{[d_1, \ldots, d_j]}(w_{j+1}, \ldots, w_k)$$
over all tuples $(d_1, \ldots, d_j) \in D^j$, where
$R_{[d_1, \ldots, d_j]}$ denotes the relation
$\{ (d_{j+1}, \ldots, d_k): (d_1, \ldots, d_k) \in R \}$.
That is, we create an extended constraint for every possible instantiation
to the universally quantified variables of the constraint.
With this observation, we can see that
all constraints can be converted to extended constraints 
in an instance of the QCSP, in polynomial time.
We can thus conclude that, as with the QCSP,
 quantified formulas with extended constraints
are PSPACE-complete in general.


We now give the official definitions for our new model.

\begin{definition}
The $\qcspe(\Gamma)$ problem is to decide the truth of a quantified formula
$\exists X_1 \forall Y_1 \exists X_2 \forall Y_2 \ldots \exists X_t \cons$
where $\cons$ is a conjunction of extended constraints, each of which
has relation from $\Gamma$.
\end{definition}

\begin{definition}
For $t \geq 1$, 
the $\tqcspe(\Gamma)$ problem is the restriction of the
$\qcspe(\Gamma)$ problem to instances having $t$ or fewer 
non-empty quantifier blocks.
\end{definition}

\begin{obs}
For all constraint languages $\Gamma$ and $t \geq 1$,
the problem $\tqcspe(\Gamma)$
is 
in the complexity class $\sigmap_{t}$ if $t$ is odd, and
in the complexity class $\pip_{t}$ if $t$ is even.
\end{obs}

\begin{obs}
For all constraint languages $\Gamma$,
the problem $\qcspe_1(\Gamma)$ is equivalent to the problem $\csp(\Gamma)$.
\end{obs}

The reader may ask why we did not define $\qcspe(\Gamma)$ in terms of
quantified formulas with \emph{constraints}: we could have
defined $\qcspe(\Gamma)$ as the problem of deciding 
those quantified formulas
$\phi$
with constraints such that 
\emph{after} the above conversion
process is applied to $\phi$ to obtain a formula $\phi'$ with
extended constraints,
each of the resulting extended constraints in $\phi'$ has
relation from $\Gamma$.
The main reason that we chose the given definition is that
we are most interested in positive results, and any positive result
concerning the model as we have defined it implies a positive result
on the alternative definition; this is because converting constraints
to extended constraints in a quantified formula can be carried out
in polynomial time, as noted above.  
We also believe that the definition of our model is very robust,
and cite the connections with the standard model developed in
Section~\ref{sect:set-functions} as evidence for this.

In this paper, we will use reductions to study the complexity of 
the problems we have defined.
We say that a problem 
\emph{reduces} to another problem if there
is a many-one polynomial-time reduction from
the first problem to the second.
We say that a class of quantified formulas \emph{uniformly reduces}
to another class of quantified formulas if
the first class reduces to the second via a reduction 
that does not increase the number of 
non-empty quantifier blocks.
Hence, if for instance
$\qcspe(\Gamma_1)$ uniformly reduces to $\qcspe(\Gamma_2)$,
then for all $t \geq 1$,
$\tqcspe(\Gamma_1)$ many-one polynomial-time reduces to $\tqcspe(\Gamma_2)$.

\subsection{Polymorphisms}
We now indicate how the algebraic, polymorphism-based 
approach that has been used to study
$\csp(\Gamma)$ and $\qcsp(\Gamma)$ complexity can be used to study
$\qcspe(\Gamma)$ complexity.  We adapt this algebraic approach in a
straightforward way, and refer the reader to~\cite{jcg97,jeavons98}
for more information on this approach.

The first point we wish to highlight is that,
 up to some mild assumptions, the set of relations
\emph{expressible} by a constraint language $\Gamma$
characterizes the complexity of $\qcspe(\Gamma)$.

\begin{definition} (see \cite{jeavons98} for details)
When $\Gamma$ is a constraint language over $D$, define 
$\lexpr \Gamma \rexpr$, the set of relations \emph{expressible} by $\Gamma$,
to be the smallest set of relations containing $\Gamma \cup \{ =_D \}$
and closed under permutation, extension, truncation, and intersection.
(Here, $=_D$ denotes the equality relation on $D$.)
\end{definition}

\begin{prop}
\label{prop:expr-complexity}
Let $\Gamma_1, \Gamma_2$ be constraint languages (over $D$)
where $\Gamma_1$ is finite and $\Gamma_2$ contains $=_D$.
If $\lexpr \Gamma_1 \rexpr \subseteq \lexpr \Gamma_2 \rexpr$, then
$\qcspe(\Gamma_1)$ uniformly reduces to $\qcspe(\Gamma_2)$.\footnote{
We remark that this proposition is not true if one removes the
assumption that $\Gamma_2$ contains $=_D$, assuming that P does not equal
NP.
Let $|D| > 1$,
let $\Gamma_2$ be the set of all constant relations,
and let $\Gamma_1$ be equal to $\Gamma_2 \cup \{=_D\}$.
Then, we have
that
 $\lexpr \Gamma_1 \rexpr \subseteq \lexpr \Gamma_2 \rexpr$,
$\qcspe(\Gamma_1)$ is coNP-hard (see Example \ref{ex:critical}), 
and $\qcspe(\Gamma_2)$ is in P.
}
(Intuitively,
the more relations that a constraint language $\Gamma$ can express,
the higher in complexity it is.)
\end{prop}

\begin{pf}
If $\lexpr \Gamma_1 \rexpr \subseteq \lexpr \Gamma_2 \rexpr$
and $\Gamma_2$ contains $=_D$, then
every constraint over $\Gamma_1$ is equivalent to 
a formula consisting of existentially quantified variables and
a conjunction of constraints over $\Gamma_2$
(see for instance the discussion in \cite{bkj01}).
Let $\phi$ be an instance of $\qcspe(\Gamma_1)$.
We create an instance of
$\qcspe(\Gamma_2)$ from $\phi$ as follows.
For each extended constraint
$$(y_1 = d_1) \wedge \ldots \wedge (y_m = d_m) \Rightarrow R(x_1,\ldots,x_k)$$
in $\phi$, let
$$\exists w_1 \ldots \exists w_n (T_1(v^1_1, \ldots, v^1_{k_1}) \wedge \ldots \wedge T_p(v^p_1, \ldots, v^p_{k_p}))$$
 be a formula 
that is equivalent to $R(x_1, \ldots, x_k)$ and where the $T_i$
are contained in $\Gamma_2$.
Replace the extended constraint
$$(y_1 = d_1) \wedge \ldots \wedge (y_m = d_m) \Rightarrow R(x_1,\ldots,x_k)$$
by the $p$ extended constraints of the form
$$(y_1 = d_1) \wedge \ldots \wedge (y_m = d_m) \Rightarrow T_i(v^i_1, \ldots, v^i_{k_i})$$
and add the variables $\{ w_1, \ldots, w_n \}$
to the innermost block of existentially quantified variables.
\end{pf}

From Proposition~\ref{prop:expr-complexity}, we can see that when
investigating finite constraint languages containing the equality relation,
any two such constraint languages
 expressing exactly the same relations
are uniformly reducible to one another, and hence of the same 
complexity in both the $\qcspe(\Gamma)$ and $\tqcspe(\Gamma)$ frameworks.

\begin{definition}
An operation $\mu: D^k \rightarrow D$ is a \emph{polymorphism}
of a relation $R \subseteq D^m$ if for any choice of $k$ tuples
$$(t^1_1, \ldots, t^1_m), \ldots, (t^k_1, \ldots, t^k_m) \in R,$$
the tuple 
$(\mu(t^1_{1}, \ldots, t^k_{1}), \ldots, 
  \mu(t^1_{m}, \ldots, t^k_{m}))$
is in $R$.
When an operation $\mu$ is a polymorphism of a relation $R$, we also say that
$R$ is invariant under $\mu$.
An operation $\mu$ is a polymorphism of a constraint language $\Gamma$
if $\mu$ is a polymorphism of all relations $R \in \Gamma$.
\end{definition}

We will be interested in the set of all polymorphisms of a constraint
language $\Gamma$, as well as the set of all relations invariant
under all operations in a given set.

\begin{definition}
Let $\od$ denote the set of all finitary operations on $D$, and let
$\rd$ denote the set of all finite arity relations on $D$.
When $\Gamma \subseteq \rd$ 
is a constraint language,
we define
$$\pol(\Gamma) = \{ \mu \in \od ~|~\mu \mbox{ is a polymorphism of } \Gamma \}.$$
When $F$ is a set of operations over $D$,
we define
$$\inv(F) = \{ R \in \rd ~|~R \mbox{ is invariant under all } \mu \in F \}.$$
\end{definition}

It is known that the expressive power of a constraint language is
determined by its polymorphisms, that is,
 $\lexpr \Gamma \rexpr = \inv(\pol(\Gamma))$.
Consequently, the complexity of a constraint language is determined
by its polymorphisms, and we have the following analog of
Proposition~\ref{prop:expr-complexity}.

\begin{prop}
\label{prop:pol-complexity}
Let $\Gamma_1, \Gamma_2$ be constraint languages (over $D$)
where $\Gamma_1$ is finite and $\Gamma_2$ contains $=_D$.
If $\pol(\Gamma_2) \subseteq \pol(\Gamma_1)$, then
$\qcspe(\Gamma_1)$ uniformly reduces to $\qcspe(\Gamma_2)$.
\end{prop}

In light of the above discussion, it makes sense to
directly define $\qcspe(F)$ for a set of operations $F$:
we define $\qcspe(F)$ as the problem $\qcspe(\inv(F))$,
and we define $\tqcspe(F)$ analogously.

From Proposition~\ref{prop:pol-complexity}, it can be seen that 
constraint languages having ``many'' polymorphisms are easier than
those having ``fewer''.  Correspondingly, many of the results on
$\csp(\Gamma)$ complexity show that the presence of a certain type
of polymorphism implies polynomial-time decidability.  
The positive complexity results in this paper will also have this form,
that is, we will prove results showing that if 
a constraint language $\Gamma$ has a polymorphism
of a certain type, then $\tqcspe(\Gamma)$ is in $\conp$.

\subsection{Relational structures}
It has been observed~\cite{fv98} that the CSP can be formulated 
as the \emph{homomorphism problem}
 of deciding, given a pair $(\rela,\relb)$ of
relational structures, whether or not there is a homomorphism
from $\rela$ to $\relb$.  We will make use of relational structures
 in this paper, and introduce them here.
A \emph{vocabulary} $\sigma$ is a collection of 
\emph{relation symbols}, each of which has an associated arity.
A \emph{relational structure} $\rela$ (over vocabulary $\sigma$)
consists of a \emph{universe} $A$, 
which is a set of size greater than or equal to one, 
and a relation $R^{\rela} \subseteq A^k$ 
for each relation symbol $R$ of $\sigma$, where $k$ is the arity 
associated to $R$.  
In this paper, we only consider relational structures having finite-size
universes.
When $\rela$ and $\relb$ are relational structures
over the same vocabulary $\sigma$, a \emph{homomorphism} from $\rela$
to $\relb$ is a mapping $h$ from the universe of $\rela$ 
to the universe of $\relb$ such that for
every relation symbol $R$ of $\sigma$ and every tuple
$(a_1, \ldots, a_k) \in R^{\rela}$, it holds that
$(h(a_1), \ldots, h(a_k)) \in R^{\relb}$.  Two relational structures
$\rela$ and $\relb$ are \emph{homomorphically equivalent} if there
is a homomorphism from $\rela$ to $\relb$ and a homomorphism from $\relb$
to $\rela$.

We say that a constraint language $\Gamma$ corresponds to
a relational structure $\relb$ (over $\sigma$) 
if each relation $S$ of $\Gamma$ can be put in a one-to-one correspondence
with a relation symbol $R$ of $\sigma$ so that
$S = R^{\relb}$.  
We define $\qcspe(\relb)$ as the problem $\qcspe(\Gamma)$ for 
the constraint language $\Gamma$ corresponding to $\relb$; and,
we use $\relbg$ to denote a relational structure corresponding to 
a constraint language $\Gamma$.
We can translate
a conjunction of constraints $\cons$ over $\Gamma$ 
to the instance $(\rela,\relbg)$
of the homomorphism problem where the universe of $\rela$ contains
all variables occurring in $\cons$ and,
for each relation symbol $R$,
the relation $R^{\rela}$ contains all tuples $(a_1, \ldots, a_k)$
such that the constraint $R^{\relbg}(a_1, \ldots, a_k)$ appears in $\cons$.

For a relational structure $\relb$ with universe $B$, 
the relational structure
$\pow(\relb)$ is defined as follows.
The universe of $\pow(\relb)$ is $\pow(B)$, where
$\pow(B)$ denotes the power set of $B$ excluding the empty set.
For every relation symbol $R$ of arity $k$
and non-empty subset $S \subseteq R^{\relb}$,
the relation $R^{\pow(\relb)}$ contains the tuple 
$(S_1, \ldots, S_k)$ where 
$S_i = \{ b_i : (b_1, \ldots, b_k) \in S  \}$.

\section{Complexity}


This section demonstrates some basic complexity properties
of our new model $\qcspe(\Gamma)$.  First, we show that
under the very mild assumption of \emph{criticality}, a constraint language
is $\conp$-hard in this model.

\begin{definition}
A constraint language $\Gamma$ is \emph{critical}
if there is an algorithm that, given a positive integer  $n \geq 2$,
outputs in time polynomial in $n$, 
sets of constraints $\cons_1, \ldots, \cons_n$ over $\Gamma$
such 
 that $\cup_{i \in \{ 1, \ldots, n \}} \cons_i$ is unsatisfiable, but 
for any $j \in \{1, \ldots, n\}$,
$\cup_{i \in \{ 1, \ldots, n \} \setminus \{ j \}} \cons_n$ is satisfiable.
\end{definition}

\begin{prop}
\label{prop:critical-hard}
If $\Gamma$ is a critical constraint language,
then for all $t \geq 2$, the problem $\tqcspe(\Gamma)$ is $\conp$-hard.
In particular, the $\forall \exists$ formulas of $\qcspe(\Gamma)$ are
$\conp$-hard.
\end{prop}

\begin{pf}
We reduce from the complement of the 3-SAT problem.
Take an instance $\phi$ of the 3-SAT problem with variables
$Y$ and clauses $C_1, \ldots, C_n$.
Compute sets of constraints $\cons_1, \ldots, \cons_n$ with the
property given in the definition of critical constraint language,
and let $X$ denote the variables occurring in the constraints $\cons_i$.

We create an instance $\phi'$ of $\qcspe(\Gamma)$ with quantifier prefix
$\forall Y \exists X$.  The extended constraints in $\phi'$ are created
as follows.
Fix two distinct elements $a_0, a_1$ of the domain of $\Gamma$.
For every clause $C_i$, 
for every constraint $R(x_1, \ldots, x_k)$ in $\cons_i$, 
and for every variable $v$ occurring in $C_i$,
create an extended constraint
$$(v = d) \Rightarrow R(x_1, \ldots, x_k)$$
where
$d$ is equal to $a_0$ if $v$ is negated in $C_i$, and equal to $a_1$
otherwise.

Observe that if
$f: Y \rightarrow \{ 0, 1 \}$ is an assignment satisfying the instance
$\phi$ of 3-SAT, 
then the created extended constraints are false
under the assignment $g: Y \rightarrow \{ a_0, a_1 \}$ defined by
$g(y) = a_{f(y)}$, and so the created instance $\phi'$ is false.
Conversely, if the instance $\phi'$ is false, then the created
extended constraints must be false under some assignment
$g: Y \rightarrow \{ a_0, a_1 \}$; in this case,
it can be verified that the
assignment $f: Y \rightarrow \{ 0, 1 \}$ defined by
$g(y) = a_{f(y)}$ satisfies all clauses of $\phi$.
\end{pf}

\begin{example}
\label{ex:critical}
Suppose $\Gamma$ is a constraint language containing the equality
relation $=_D$
and (at least) two
different constant relations $R_a, R_b$.
The constraint language $\Gamma$ is critical: for any $n \geq 2$,
the sets of constraints
$\{ =_D(v_1,v_2), R_a(v_1) \}$, $\{ R_b(v_n) \}$, and
$\{ =_D(v_i,v_{i+1}) \}$ for $i \in \{ 2, \ldots, n-1 \}$ has the
desired property, and can easily be generated in polynomial time.
\end{example}

Although our $\qcspe(\Gamma)$ model is in general $\conp$-hard, 
we can use it to obtain positive complexity results for QCSP instances
for which no complexity result can be derived in the standard model,
other than the trivial $\pspace$ upper bound.  We discuss this phenomenon
in the following examples.

\begin{example}
Let us consider \emph{extended quantified {\sc 2-SAT} formulas}, which
we define to be instances of the boolean QCSP
 where each constraint must be 
a clause
in which there are at most two 
occurrences of existential variables.
We call such a constraint
 an \emph{extended 2-clause}.
Recall that
a clause is a disjunction of literals, where a literal is a variable
or its negation.  The following are examples of extended 2-clauses:
$$\overline{y_1} \vee y_4 \vee \overline{x_1} \vee \overline{x_2}$$
$$x_1 \vee y_2 \vee \overline{x_3} \vee \overline{y_5} \vee y_8$$
Here, the $y_i$  denote universal variables, and the $x_i$ denote
existential variables.
For any tuple $(a_1, \ldots, a_k) \in \{ 0, 1 \}^k$, let
$R_{(a_1, \ldots, a_k)}$ denote the relation
$\{ 0, 1 \}^k \setminus \{ (a_1, \ldots, a_k) \}$.
Notice that each extended 2-clause is equivalent to a constraint
of the form 
$R_{(a_1, \ldots, a_k)}(v_1, \ldots, v_k)$.
For example, the two given clauses are equivalent to the constraints:
$$R_{(1,0,1,1)}(y_1,y_4,x_1,x_2)$$
$$R_{(0,0,1,1,0)}(x_1,y_2,x_3,y_5,y_8)$$
If we are to model 
extended quantified {\sc 2-SAT} formulas
using the standard model
$\qcsp(\Gamma)$, we take as our constraint language the set
of all relations that can appear in constraints, that is, 
the set of all relations $R_{(a_1, \ldots, a_k)}$.
This constraint language is easily seen to give rise to a $\pspace$-complete
case of the QCSP, under the standard model:
it can directly encode {\sc Quantified 3-SAT}.

On the other hand, we can model 
extended quantified {\sc 2-SAT} formulas
under our new model as the problem $\qcspe(\Gamma_2)$, where
$\Gamma_2$ is the constraint language 
$\{ R_{(0)}, R_{(1)}, R_{(0,0)}, R_{(0,1)}, R_{(1,1)} \}$. 
For example, the two given clauses are equivalent to the extended constraints:
$$(y_1 = 1) \wedge (y_4 = 0) \Rightarrow  R_{(1,1)}(x_1, x_2)$$
$$(y_2 = 0) \wedge (y_5 = 1) \wedge (y_8 = 0) \Rightarrow R_{(0,1)}(x_1, x_3)$$
Let $m: \{ 0, 1 \}^3 \rightarrow \{ 0, 1 \}$ be the majority operation on
$\{ 0, 1 \}$, that is, the symmetric operation that returns $0$ if two
or three of its arguments are equal to $0$, and $1$ if two or three
of its arguments are equal to $1$.  
It can be verified that the constraint language $\Gamma_2$ has $m$ as
a polymorphism.
The operation $m$ is an example of a \emph{near-unanimity} operation;
later in this paper (Theorem~\ref{thm:near-unanimity}), 
we show that for any constraint language
$\Gamma$ having a near-unanimity polymorphism, the problem
$\tqcspe(\Gamma)$ is in $\conp$ (for all $t \geq 2$).
Hence, we have in particular that $\tqcspe(\Gamma_2)$ is in $\conp$
(for all $t \geq 2$), that is,
extended quantified {\sc 2-SAT} formulas
are in $\conp$, under bounded alternation.
\end{example}

\begin{example}
\label{ex:extended-quantified-horn}
\emph{Extended quantified Horn formulas} were introduced by 
Kleine~B\"{u}ning~et~al.~\cite{bkf95}.  An extended quantified Horn formula
is an instance of the boolean QCSP where every constraint is
an \emph{extended Horn clause}, that is, a clause in which there 
is at most one positive literal of an existential variable.  In other words,
an extended Horn clause is a clause where removing all literals of universal
variables results in a Horn clause.

Let $H \subseteq \{ 0, 1 \}^3$ be the relation
$\{ 0, 1 \}^3 \setminus \{ (1, 1, 0) \}$, and let $\Gamma_H$ 
be the constraint language $\{ H, R_{0}, R_{1} \}$, where
$R_{0}$ and $R_{1}$ are defined as the constant relations
$\{ (0) \}$ and $\{ (1) \}$, respectively.
A given extended quantified Horn formula can easily be translated
(in polynomial time) into an instance of the problem $\qcspe(\Gamma_H)$.
In particular, any extended Horn clause can be translated into an
existentially quantified
conjunction of extended constraints over $\Gamma_H$,
and so the idea of the proof of Theorem~\ref{prop:expr-complexity}
can be applied.
For example, consider the following two extended Horn clauses:
$$x_1 \vee \overline{x_2} \vee y_1 \vee \overline{y_2} \vee \overline{x_3} \vee \overline{x_4}$$
$$\overline{y_1} \vee \overline{x_1} \vee x_2$$
They are equivalent to the following existentially quantified formulas
over $\Gamma_H$:
\begin{center}
$\exists v [ ((y_1 = 0) \wedge (y_2 = 1) \Rightarrow H(x_2, x_3, v)) \wedge$\\
$((y_1 = 0) \wedge (y_2 = 1) \Rightarrow H(v, x_4, x_1))]$
\end{center}
$$\exists v[ (R_{1}(v)) \wedge ((y_1 = 1) \Rightarrow H(v, x_1, x_2)) ]$$
The boolean AND operation is a polymorphism of the constraint language
$\Gamma_H$, and is an example of a semilattice operation; 
later in the paper (Corollary~\ref{cor:semilattice}),
we show that
for any constraint language
$\Gamma$ having a semilattice polymorphism, the problem
$\tqcspe(\Gamma)$ is in $\conp$ (for all $t \geq 2$).
This implies that extended quantified Horn formulas are in $\conp$,
under bounded alternation; this is the first non-trivial
complexity upper bound on this class of formulas.
\end{example}

We observe that our model is at least as hard as the standard model,
with respect to constraint languages containing all constants.

\begin{prop} 
\label{prop:standard-to-our}
If $\Gamma$ be a constraint language containing all constant relations,
then $\qcsp(\Gamma)$ uniformly reduces to $\qcspe(\Gamma)$.
\end{prop}

\begin{pf}
Given an instance $\phi$ of $\qcsp(\Gamma)$, we create an instance $\phi'$
of $\qcspe(\Gamma)$ as follows.  The quantifier prefix of $\phi'$ is
equal to that of $\phi$, except there are $|D|$ extra existentially
quantified variables (which may be existentially quantified anywhere),
denoted by $\{ v_d: d \in D \}$.  The instance $\phi'$ contains
the extended constraints $\{ R_d(v_d): d \in D \}$,
where $R_d$ denotes the constant relation $\{ (d) \}$;
these extended constraints ``force''
each variable $v_d$ to take on the value $d$.
For each constraint $C$ of $\phi$, we also create extended constraints
in $\phi'$, as follows.  For the sake of notation, let us denote
$C$ by $R(y_1, \ldots, y_m, x_1, \ldots, x_n)$ where the $y_i$
are universally quantified and the $x_i$ are existentially quantified.
For each tuple $(d_1, \ldots, d_m) \in D^m$, create an extended constraint
$$(y_1 = d_1) \wedge \ldots \wedge (y_m = d_m) \Rightarrow 
R(v_{d_1}, \ldots, v_{d_m}, x_1, \ldots, x_n).$$
\end{pf}

We now turn to look at some algebraic properties of our new model.

\begin{theorem}
\label{thm:hom-equivalent}
If $\relb$, $\relb'$ are homomorphically equivalent relational structures
(over a finite vocabulary),
then $\qcspe(\relb)$ and $\qcspe(\relb')$
uniformly reduce to each other.
\end{theorem}

\begin{pf}
First, suppose that $\relc$ and $\relc'$ are relational structures
such that $\relc$ has universe $\{ c \}$ of size one, 
and there is a homomorphism $h$ from $\relc$ to $\relc'$.
Then, every relation of $\relc'$ is either empty or contains
the tuple $(h(c), \ldots, h(c))$ with all coordinates equal to $h(c)$.
It follows that $\qcsp(\relc')$ can be easily decided in polynomial time:
an instance is true as long as there
is no extended constraint
$(y_1 = d_1) \wedge \ldots \wedge (y_m = d_m) \Rightarrow R(x_1,\ldots,x_k)$
such that $R$ is empty and there is an assignment mapping
$y_i$ to $d_i$ for all $i$.
(Likewise, every relation of $\relc$ is either empty or contains
the tuple $(c, \ldots, c)$ with all coordinates equal to $c$;
and $\qcsp(\relc)$ can be easily decided in polynomial time.)
We thus assume that both $\relb$ and $\relb'$ has universe of size
strictly greater than one.

We show how to reduce from $\qcspe(\relb)$ to $\qcspe(\relb')$.
Let $h$ be a homomorphism from $\relb$ to $\relb'$, 
let $h'$ be a homomorphism from $\relb'$ to $\relb$, and
let $\phi$ be an arbitrary instance of $\qcspe(\relb)$.
We create an instance of $\qcspe(\relb')$ as follows.
First, let $\phi'$ be the quantified formula obtained from $\phi$
by replacing each relation $R^{\relb}$ with $R^{\relb'}$.
Letting it be understood that the universally quantified variables of $\phi'$
are still quantified over the universe $B$ of $\relb$, it can be
verified that 
if $\{ \sigma_x \}$ is a winning strategy for
$\phi$, then $\{ h\sigma_x \}$ is a winning strategy for $\phi'$; 
and likewise,
that
if $\{ \sigma'_x \}$ is a winning strategy for
$\phi'$, then $\{ h'\sigma'_x \}$ is a winning strategy for $\phi$.
Now, it remains to modify $\phi'$ so that the universally quantified
variables are quantified over the universe $B'$ of $\relb'$.
If $|B'| \geq |B|$, it suffices to take any injective mapping
$i: B \rightarrow B'$ and simply replace all instances of $b$ 
in $\phi'$ by $i(b)$.  If $|B'| < |B|$, then each universally
quantified variable over $B$ can be simulated by $s$ 
universally quantified variables over $B'$, where
$s$ is a sufficiently large constant so that $|B'|^s \geq |B|$.  Notice
that such a constant $s$ exists since $|B'| \geq 2$.
\end{pf}

When $\alga$ is an algebra with operations $F$, we define
$\qcspe(\alga)$ as $\qcspe(F)$.

\begin{theorem}
\label{thm:sub-alg-hom}
Let $\alga$ be a finite algebra.
\begin{itemize}
\item If $\algb$ is a subalgebra of $\alga$, then 
$\qcspe(\algb)$ uniformly reduces to $\qcspe(\alga)$.

\item If $\algb$ is a homomorphic image of $\alga$, then 
$\qcspe(\algb)$ uniformly reduces to $\qcspe(\alga)$.

\end{itemize}
\end{theorem}

\begin{pf}
The proof of this theorem follows proofs of similar results 
in~\cite{bkj00}.  If $\algb$ is a subalgebra of $\alga$,
then $\qcspe(\algb)$ reduces to $\qcspe(\alga)$ by the identity mapping.
Now let $\algb$ be a homomorphic image of $\alga$, and let
$f$ be a surjective homomorphism from $\alga$ to $\algb$.
Let $\phi$ be an instance of $\qcspe(\algb)$, and let
$\relb_\algb$ the relational structure whose relations are exactly
the relations occurring in $\phi$.
Define a relational structure $\relb_\alga$ over the same vocabulary
as $\relb_\algb$ where
for each relation symbol $R$, the relation
$R^{\relb_\alga}$ is defined as 
$\{ (a_1, \ldots, a_k): (f(a_1), \ldots, f(a_k)) \in R^{\relb_\algb} \}$.
All relations $R^{\relb_\alga}$ are invariant under the operations
of $\alga$.
Let $f'$ be any mapping from the universe $B$ of $\algb$ to the
universe $A$ of $\alga$ such that $f'(b) \in f^{-1}(b)$.
Then, 
$f$ is a homomorphism from $\relb_\alga$ to $\relb_\algb$, 
$f'$ is a homomorphism from $\relb_\algb$ to $\relb_\alga$, 
and the reduction of Theorem~\ref{thm:hom-equivalent}
can be employed.
\end{pf}

\section{Fingerprints}
\label{sect:fingerprints}

\newcommand{\arity}{\mathsf{arity}}
\newcommand{\rel}{\mathcal{R}}
\newcommand{\pr}{\mathsf{pr}}
\newcommand{\lf}{\langle}
\newcommand{\rf}{\rangle}

\newcommand{\lphi}{<^{\phi}}
\newcommand{\leqphi}{\leq^{\phi}}

\newcommand{\algproj}{\mathsf{Proj}}
\newcommand{\alginf}{\mathsf{Inf}}
\newcommand{\algcons}{\mathsf{Cons}}


This section introduces the notion of a fingerprint along with 
various associated notions.
In this section and the next, we are concerned primarily with
our new existentially restricted model of quantified constraint satisfaction,
and so we
assume that all quantified
formulas under discussion contain extended constraints.

We define a projection operator $\pr_k$ which projects ``onto the
first $k$ coordinates'': formally, when $R$ is a relation of arity $n$
and $k \in \{ 0, \ldots, n \}$, we define
$\pr_k R = \{ (d_1, \ldots, d_k): (d_1, \ldots, d_n) \in R \}$.
At this point,
we adopt the convention that there is a unique tuple of arity $0$, and
hence a unique non-empty relation of arity $0$.  When a non-empty (empty)
relation is projected down to an arity $0$ relation,
we consider the result to be the
unique non-empty (respectively, empty) relation of arity $0$.

\begin{definition}
\label{fingerprint-collection}

A \emph{fingerprint collection} is a set $\fcoll$ with an associated
domain $D$ whose elements are called \emph{fingerprints}.
Each fingerprint $F \in \fcoll$ has an associated arity, denoted
by $\arity(F)$, and specifies a relation
$\rel(F) \subseteq D^{\arity(F)}$.
We require that there is a fingerprint $\top \in \fcoll$ such that
$\rel(F)$ is the non-empty relation of arity $0$.
We require that there is a projection function $\pi$ such that
\begin{itemize}
\item given any fingerprint
$F \in \fcoll$ and $k \in \{ 0, \ldots, \arity(F) \}$, outputs a fingerprint
$\pi_k F$ such that $\rel(\pi_k F) = \pr_k \rel(F)$, and

\item when $0 \leq k \leq l \leq n$ and $F \in \fcoll$ is of arity $n$,
it holds that $\pi_k F = \pi_k (\pi_l F)$.
\end{itemize}
Also, we require that there is a preorder $\sqsubseteq$ such that
\begin{itemize}
\item $F \sqsubseteq F'$ implies
$\arity(F) = \arity(F')$ and 
$\rel(F) \subseteq \rel(F')$,

\item if $F \sqsubseteq F'$ and $k \in \{ 0, \ldots, \arity(F) \}$,
then $\pi_k F \sqsubseteq \pi_k F'$, and

\item with respect to $\sqsubseteq$, chains are of polynomial length;
that is, 
there exists a polynomial $p$ such that if
$F_1, \ldots, F_m \in \fcoll$ are distinct fingerprints of arity $n$ and
$F_1 \sqsubseteq \cdots \sqsubseteq F_m$, then 
$m \leq p(n)$.
\end{itemize}
Finally, we require that
 each fingerprint $F \in \fcoll$ has a representation with size polynomial
in its arity, that is, there is a function 
$r: \fcoll \rightarrow \{0,1\}^*$ and a polynomial $q$ such that
$|r(F)| \leq q(\arity(F))$ for all $F \in \fcoll$.

\end{definition}

For ease of notation we will always denote the representation $r(F)$ of
a fingerprint $F$ simply by $F$, although technically all of the
algorithms that we will discuss
manipulate representations of fingerprints.

\begin{example}
\label{ex:collection-constant}
A simple example of a fingerprint collection is as follows.
Let $D$ be any domain, and fix $d$ to be any element of $D$.
Define $\fcoll_d = \{ \top_0, \bot_0, \top_1, \bot_1, \ldots \}$
where each $\bot_i$ has arity $i$ and has $\rel(\bot_i) = \emptyset$,
and each $\top_i$ has arity $i$ and is non-empty with
$\rel(\top_i)$ containing the unique tuple 
of arity $i$ equal to $d$ at all coordinates, that is,
$(d, \ldots, d)$.  
We can define the projection function $\pi$ by
$\pi_k \top_n = \top_k$ and $\pi_k \bot_n = \bot_k$, for $n \geq k$.
We define the preorder $\sqsubseteq$ by $\bot_i \sqsubseteq \top_i$.
Chains are clearly of length at most two.
The fingerprint collection
$\fcoll_d$ clearly admits a polynomial size representation, 
as each fingerprint $F \in \fcoll_d$ can simply be encoded by
its arity along with a bit denoting whether it is of type $\top$ or $\bot$.
\end{example}

\begin{example}
\label{ex:collection-set-function}
A perhaps more interesting example of a fingerprint collection is
as follows.  Fix a domain $D$ and let $\fcollp$ contain
all tuples $(D_1, \ldots, D_n)$ where each $D_i$ is a non-empty subset
of $D$.  
We define $\rel( (D_1, \ldots, D_n) )$ to be the set of all $n$-tuples
$(d_1, \ldots, d_n)$ such that $d_i \in D_i$ for all 
$i \in \{ 1, \ldots, n \}$.
Let $\fcollp$ also contain elements
$\bot_0, \bot_1, \ldots$ where $\bot_i$ is of arity $i$ with
$\rel(\bot_i) = \emptyset$.
We can define the projection function $\pi$ by
$\pi_k (D_1, \ldots, D_n) = (D_1, \ldots, D_k)$ and
$\pi_k \bot_n = \bot_k$, for $n \geq k$.

We define the preorder $\sqsubseteq$ by the following rule:
$(D_1, \ldots, D_n) \sqsubseteq (D'_1, \ldots, D'_n)$ if and only if
$D_i \subseteq D'_i$ for all $i \in \{ 1, \ldots, n \}$.  Notice that
if $F_1, \ldots, F_m \in \fcollp$ are distinct fingerprints of arity $n$ and
$F_1 \sqsubseteq \cdots \sqsubseteq F_m$, 
then $m \leq n |D|$, and so chains are of linear length.
(Recall that all domains $D$ in this paper are finite).

It is straightforward to give a polynomial (in fact, linear)
size representation for the fingerprints
of $\fcollp$, as there
 are a constant number of subsets of $D$.
\end{example}

\begin{definition}
A \emph{fingerprint application} is a fingerprint $F$ paired with a
tuple of variables $\lf x_1, \ldots, x_k \rf$ of length $k = \arity(F)$,
and is denoted by $F\lf x_1, \ldots, x_k \rf$.
It is considered to be true under an assignment $f$ defined on 
$\{ x_1, \ldots, x_k \}$
if $(f(x_1), \ldots, f(x_k)) \in \rel(F)$.
\end{definition}

While a fingerprint application is similar to a constraint, we use
the ``angle brackets'' notation for fingerprint applications to 
differentiate between the two.

\begin{definition}
\label{def:fingerprint-scheme}
A \emph{fingerprint scheme} for a constraint language $\Gamma$ consists of:
\begin{itemize}

\item A fingerprint collection $\fcoll$ over the domain $D$ of $\Gamma$.

\item A \emph{projection algorithm} running in polynomial time
that, given a fingerprint $F \in \fcoll$ of arity $n$ and
an integer $k \in \{ 0, \ldots, n \}$, computes the fingerprint
$\pi_k F$,


\item An \emph{inference algorithm} $\alginf$ running in polynomial time
that, given 
\begin{itemize}
\item a fingerprint application $F\lf x_1, \ldots, x_k \rf$, and

\item a conjunction of constraints $\cons$ over $\Gamma$ and variables 
$\{ x_1, \ldots, x_n \}$, where $n \geq k$,
\end{itemize}
computes a fingerprint
$F'$ of arity $n$ where 
\begin{enumerate}
\item (soundness)
for all assignments $f: \{ x_1, \ldots, x_n \} \rightarrow D$, if both 
$F\lf x_1, \ldots, x_k \rf$ and $\cons$ are true under $f$, then
$F'\lf x_1, \ldots, x_n \rf$ is also true under $f$, and 
\item (progress) it holds that $\pi_{k} F' \sqsubseteq F$.
\end{enumerate}

Under the two given assumptions, 
we say that 
$F'\lf x_1, \ldots, x_n \rf$ is a fingerprint application 
\emph{suitable} for $\cons$
if there exists
a fingerprint application $F\lf x_1, \ldots, x_k \rf$ such that
$\alginf( F\lf x_1, \ldots, x_k \rf, \cons ) = F'$.

\item A \emph{construction mapping} $\algcons: \fcoll \rightarrow D$ 
such that 
when $F\lf x_1, \ldots, x_n \rf$ 
is a fingerprint application suitable for 
a conjunction of constraints $\cons$ 
(over $\Gamma$ and variables $\{ x_1, \ldots,  x_n \}$ with $n \geq 1$)
and $\rel(F) \neq \emptyset$, the mapping taking
$x_i$ to $\algcons(\pi_i F)$, for all $i \in \{ 1, \ldots, n \}$,
satisfies $\cons$.
\end{itemize}
\end{definition}

\begin{example}
We give an example of a fingerprint scheme that uses the fingerprint collection
of
Example \ref{ex:collection-constant}.
Let $D$ be any domain and $d \in D$ be an element of $D$.
Suppose that $\Gamma$ is a constraint language over $D$ that is
\emph{$d$-valid} in that every 
non-empty relation $R \in \Gamma$
contains the all-$d$ tuple $(d, \ldots, d)$ having the arity of $R$.
We demonstrate that there is a fingerprint scheme for $\Gamma$.
The fingerprint collection is $\fcoll_d$, defined in 
Example \ref{ex:collection-constant}.
It is clear that projections can be computed in polynomial time.
The inference algorithm, 
given $F\lf x_1, \ldots, x_k \rf$ and a conjunction of constraints
$\cons$ over $\Gamma$ and variables $\{ x_1, \ldots, x_n \}$, 
outputs $\bot_{n}$ if $F = \bot_{k}$ or $\cons$ contains a constraint
with empty relation, and outputs $\top_{n}$ otherwise.
It is straightforward to verify that this inference algorithm obeys the 
soundness and progress conditions.
The construction mapping for our fingerprint scheme simply always outputs
$d$.  This mapping satisfies the requirement for a construction
mapping: if $F$ is a fingerprint with $\rel(F) \neq \emptyset$, 
then $F = \top_n$ (where $n$ is the
arity of $F$); when $\top_n\lf x_1, \ldots, x_n \rf$ 
is a fingerprint application
suitable for constraints $\cons$ over $\Gamma$ and $\{ x_1, \ldots, x_n \}$, 
no constraint in $\cons$ may have
empty relation, so every constraint in $\cons$ is $d$-valid,
and thus the assignment mapping every $x_i$ to $d$
satisfies $\cons$.
\end{example}


\begin{example}
\label{ex:scheme-set-function}
We continue Example \ref{ex:collection-set-function} by giving a fingerprint
scheme using the fingerprint collection defined there.
This example makes use of ideas concerning set functions in the context of
constraint satisfaction as well as the notion of
\emph{arc consistency}; 
for more information, we refer the readers to
the papers
\cite{dp99,cd04-slaac}.
We consider a \emph{set function} to be a mapping 
$f: \pow(D) \rightarrow D$, where $\pow(D)$ denotes the 
power set of $D$ excluding the empty set.
We say that $f$ is a polymorphism of
a constraint language $\Gamma$ 
if all of the functions $f_i: D^i \rightarrow D$ defined by
$f_i(x_1, \ldots, x_i) =
 f(\{x_1, \ldots, x_i\})$, for $i \geq 1$,
 are polymorphisms of $\Gamma$.
Equivalently, $f$ is a polymorphism of $\Gamma$ if 
$f$ is a homomorphism from $\pow(\relbg)$ to $\relbg$.

Let $\Gamma$ be a constraint language over domain $D$ having
a set function
$f: \pow(D) \rightarrow D$ as polymorphism.
We demonstrate a fingerprint scheme for $\Gamma$.
The fingerprint collection is $\fcollp$, from Example
\ref{ex:collection-set-function}.  Projections of fingerprints
can clearly be computed in polynomial time.
The inference algorithm is \emph{arc consistency}.
More specifically,
the inference algorithm takes as input
a fingerprint application $(D_1, \ldots, D_k) \lf x_1, \ldots, x_k \rf$ 
and a conjunction of constraints
$(\rela,\relbg)$ over $\Gamma$ and variables $\{ x_1, \ldots, x_n \}$.
The algorithm tries to establish arc consistency on
$(\rela,\relbg)$ with additional constraints stating that $x_i \in D_i$
for all $i \in \{ 1, \ldots, k \}$.
The result is either
 a homomorphism
$g: \rela \rightarrow \pow(\relbg)$
with $g(x_i) \subseteq D_i$ for all $i \in \{ 1, \ldots, n \}$
such that the fingerprint $(g(x_1), \ldots, g(x_n))$
satisfies the soundness and progress requirements;
or, certification that arc consistency cannot be established,
in which case the fingerprint $(\emptyset, \ldots, \emptyset)$ 
satisfies the soundness and progress requirements.

The construction mapping is defined by
$\algcons((D_1, \ldots, D_n)) = f(D_n)$.
To see that this algorithm satisfies the given criterion,
assume that $(D_1, \ldots, D_n)\lf x_1, \ldots, x_n \rf$ 
is a fingerprint application
suitable for a conjunction of constraints 
$(\rela,\relbg)$.
The mapping $h: \{ x_1, \ldots, x_n \} \rightarrow D$ given by
the construction mapping obeys $h(x_i) = f(D_i)$.
By assumption, the mapping taking $x_i$ to $D_i$ is a homomorphism
$g: \rela \rightarrow \pow(\relbg)$.
Observe that $h$ is the composition
of
$g: \rela \rightarrow \pow(\relbg)$
and
$f: \pow(\relbg) \rightarrow \relbg$,
 and so $h$ is a homomorphism from
$\rela$ to $\relbg$, as desired.
\end{example}

\section{Proof System}
\label{sect:proof-system}

\newcommand{\Rule}[2]{          
  \begin{array}{c}
  #1 \\\hline
  #2
  \end{array}}

This section presents the proof system that will permit us to derive
$\conp$-inclusion results for constraint languages having fingerprint schemes.
The proof system gives rules for deriving, from a quantified formula
$\phi$ and a fingerprint application for $\phi$, further fingerprint 
applications for $\phi$.

Before giving the proof system, we require some notation.
We assume that, for every quantified formula
$\phi = \exists X_1 \forall Y_1 \exists X_2 \forall Y_2 \ldots \exists X_t \cons$, 
there is an associated total order $\leqphi$ on the existential variables
$\cup_{i=1}^t X_i$ that respects the quantifier prefix in that
$x \leqphi x'$ if $x \in X_i$, $x' \in X_j$, and $i < j$.
We say that a tuple of variables $\lf x_1, \ldots, x_k \rf$ is a 
\emph{prefix} of $\phi$ if 
it is an ``initial segment'' of the existential variables of $\phi$
under the $\leqphi$ ordering, that is,
all $x_i$ are existentially quantified in
$\phi$, 
$x_i \leqphi x_{i+1}$ for all
$i \in \{ 1, \ldots, k-1 \}$, 
and $x \leqphi x_j$ implies that $x = x_i$ for some $i \leq j$.
When $\lf x_1, \ldots, x_k \rf$ is a prefix of a quantified formula
$\phi$ that is understood from the context and
$X = \{ x_1, \ldots, x_k\}$, we use the ``set notation'' $\lf X \rf$
to denote 
$\lf x_1, \ldots, x_k \rf$.

We now give the proof system, which consists of rules for deriving
expressions of the form
$\phi, F\lf X \rf \vdash F'\lf X' \rf$.
In such expressions, $\phi$ is always a quantified formula,
$F\lf X \rf$ and $F'\lf X' \rf$ are fingerprint applications,
and it is always assumed that
$X\subseteq X_1^{\phi}$.  Moreover, note that if
$\phi, F\lf X\rf \vdash F'\lf X' \rf$, it will always hold that
$X' = X_1^{\phi}$.

\begin{definition}
The proof system for a fingerprint scheme
with fingerprint collection $\fcoll$ and inference algorithm $\alginf$
consists of the following three rules.

$\Rule{}{(\exists X' \cons), F\lf X \rf \vdash \alginf(F\lf X \rf, \cons)\lf X' \rf}$

\vspace{5pt}

$\Rule{\phi, F\lf X \rf \vdash F'\lf X' \rf        \qquad   
       \phi, F'\lf X' \rf \vdash F''\lf X'' \rf   }
      {\phi, F\lf X   \rf \vdash F''\lf X'' \rf   }$

\vspace{5pt}

$\Rule{\phi[g], F \lf X \rf  \vdash  F' \lf X' \rf  \qquad  
                                            X' \supseteq X_1^{\phi}}
      {\phi,    F \lf X \rf  \vdash 
                    (\pi_{|X_1^{\phi}|} F')\lf X_1^{\phi} \rf}$

\vspace{5pt}

The last rule is applicable when $\phi$ has more than one existential block,
and
$g: Y_1^{\phi} \rightarrow D$ is an assignment to the first universal block.
The formula $\phi[g]$ is defined to be the formula derived from $\phi$
by removing $\forall Y_1^{\phi}$ from the quantifier prefix and instantiating
each variable occurrence $y \in Y_1^{\phi}$ in $\cons$ 
with the constant $g(y)$.
\end{definition}

\begin{prop}
\label{prop:sound}
The proof system for any fingerprint scheme is sound in the following sense:
for any quantified formula $\phi = \exists X_1 \ldots \exists X_t \cons$
and any fingerprint application $F\lf X \rf$ for $\phi$, 
if it holds that $\phi, F\lf X\rf \vdash F'\lf X'\rf$, then 
the formulas
$\exists X_1 \ldots \exists X_t [\cons \wedge F\lf X\rf]$ and 
$\exists X_1 \ldots \exists X_t [\cons \wedge F\lf X\rf \wedge F'\lf X'\rf]$
have the same winning strategies
(and hence the same truth value).
\end{prop}

\begin{pf}
Straightforward.  Note that the soundness of the
first proof rule relies on the soundness of the inference algorithm
$\alginf$.
\end{pf}

Let us say that \emph{$\phi$ has a proof of falsity} if
 $\phi, \top \vdash \bot$ for a fingerprint application 
$\bot = F'\lf X' \rf$ 
with $\rel(F') = \emptyset$.
The previous proposition implies that if a formula $\phi$ has a proof
of falsity,
then $\phi$ is indeed false.
Our next theorem implies that, when a fingerprint scheme for
a constraint language $\Gamma$ exists, the proof system for this
scheme is \emph{complete} for the class of formulas $\qcspe(\Gamma)$
in that \emph{all} false formulas have proofs of falsity.
It in fact demonstrates that for false alternation-bounded instances,
there are \emph{polynomial-size} proofs of falsity.

\begin{theorem}
\label{thm:polysize-proofs}
Suppose that $\Gamma$ is a constraint language having a fingerprint scheme.
In the above proof system,
the false formulas of $\tqcspe(\Gamma)$
have polynomial-size proofs of falsity 
(for each $t \geq 1$).
\end{theorem}

Theorem~\ref{thm:polysize-proofs} is proved in the appendix.
We derive the following consequence, the principal result of this section,
from 
Theorem \ref{thm:polysize-proofs}.

\begin{theorem}
\label{thm:in-conp}
If $\Gamma$ is a constraint language having a fingerprint scheme,
then 
the problem $\tqcspe(\Gamma)$ is in $\conp$
(for each $t \geq 2$).
\end{theorem}

\begin{pf}
Observe that
proofs 
in the above proof system can be verified in polynomial time;
in particular, instances of the third proof rule can be verified
in polynomial time as a fingerprint scheme is required to have a 
 polynomial-time projection algorithm.
The theorem is immediate from
Proposition \ref{prop:sound}, 
Theorem \ref{thm:polysize-proofs},
and this observation.
\end{pf}

\section{Applications}
\label{sect:applications}

The previous section gave a proof system for constraint languages having
fingerprint schemes, and moreover demonstrated that the given proof system
implies $\conp$ upper bounds on the complexity of such
constraint languages, in our new model.  In this section, 
we derive a number of $\conp$ upper bounds by demonstrating that various
classes of constraint languages have fingerprint schemes.
All of these classes have been previously studied in the $\csp(\Gamma)$
model, see~\cite{dp99,jcg97,jcc98,bulatov02-maltsev,dalmau04-maltsev}.
If for all $t \geq 2$ it holds that the problem $\tqcspe(\Gamma)$
is in $\conp$, we will simply say that 
the problem $\tqcspe(\Gamma)$ is in $\conp$.

\begin{theorem}
\label{thm:set-function}
If $\Gamma$ is a constraint language having a set function polymorphism,
then 
the problem $\tqcspe(\Gamma)$ is in $\conp$.
\end{theorem}

\begin{pf}
If $\Gamma$ is a constraint language having a set function polymorphism,
then it has a fingerprint scheme by the discussion in 
Examples~\ref{ex:collection-set-function} and
\ref{ex:scheme-set-function}.
The theorem thus follows from Theorem \ref{thm:in-conp}.
\end{pf}

From Theorem~\ref{thm:set-function}, we can readily derive a 
$\conp$ bound on constraint languages having a \emph{semilattice polymorphism}.
A \emph{semilattice operation} is a binary operation that is
associative, commutative, and idempotent.

\begin{corollary}
\label{cor:semilattice}
If $\Gamma$ is a constraint language having a semilattice polymorphism,
then
the problem $\tqcspe(\Gamma)$ is in $\conp$.
\end{corollary}

\begin{pf}
Suppose that $\Gamma$
is a constraint language over domain $D$
having a semilattice polymorphism $\oplus: D^2 \rightarrow D$.
Let $\Gamma'$ be defined as the set containing all relations in
$\Gamma$ as well as all constant relations of $D$.
Since $\Gamma \subseteq \Gamma'$, it suffices to show the result for
$\Gamma'$.  Observe that $\Gamma'$ is also invariant under
$\oplus$: all constant relations are preserved by $\oplus$ as
$\oplus$ is idempotent.  
It is known that any constraint language having a semilattice polymorphism
also has a set function polymorphism~\cite{dp99}, and hence
 $\tqcspe(\Gamma')$ is in $\conp$
by Theorem~\ref{thm:set-function}.
\end{pf}

A \emph{near-unanimity operation} is an 
idempotent operation $f: D^k \rightarrow D$ with
$k \geq 3$ and such that when all but one of its arguments are equal to 
an element $d \in D$,
then $f$ returns $d$;
that is, it holds that $f(a, b, \ldots, b) = f(b, a, b, \ldots, b) = \cdots = f(b, \ldots, b, a) = b$ for all $a, b \in D$.

\begin{theorem}
\label{thm:near-unanimity}
If $\Gamma$ is a constraint language having a near-unanimity polymorphism,
then 
the problem $\tqcspe(\Gamma)$ is in $\conp$.
\end{theorem}

A Mal'tsev operation is an operation $f: D^3 \rightarrow D$
such that $f(a, b, b) = f(b, b, a) = a$ for all $a, b \in D$.

\begin{theorem}
\label{thm:maltsev}
If $\Gamma$ is a constraint language having a Mal'tsev polymorphism,
then 
the problem $\tqcspe(\Gamma)$ is in $\conp$.
\end{theorem}

Theorems~\ref{thm:near-unanimity} and~\ref{thm:maltsev} are proved in the
appendix.

Using the results in this section thus far, we can readily obtain
a complexity classification theorem for 
 constraint languages over a two-element domain,
in our new model.
Let us say that the problem $\qcspe(\Gamma)$ has \emph{maximal complexity} if 
$\qcspe(\Gamma)$ is $\pspace$-complete and
for all $t \geq 1$, the problem $\tqcspe(\Gamma)$ 
is $\sigmap_t$-complete for odd~$t$, and $\pip_t$-complete for even~$t$.
(We will also apply this terminology to problems
of the form $\qcsp(\Gamma)$.)
The following theorem shows that in our model of QCSP complexity,
there is a dichotomy in the behavior of constraint languages over
a two-element domain: either they are in $\conp$ under bounded alternation,
or of maximal complexity.

\begin{theorem}
Let $\Gamma$ be a constraint language over a two-element domain $D$,
containing $=_D$.
The problem $\tqcspe(\Gamma)$
is in $\conp$ if $\Gamma$ has
\begin{itemize}
\item a constant polymorphism, 
\item a semilattice polymorphism,
\item a near-unanimity polymorphism, or 
\item a Mal'tsev polymorphism;
\end{itemize}
otherwise, $\qcspe(\Gamma)$ has maximal complexity.
\end{theorem}

\newcommand{\nae}{\mathsf{NAE}}

\begin{pf}
If $\Gamma$ has one of the four named types of polymorphisms, 
then the theorem holds by 
Corollary~\ref{cor:semilattice},
Theorem~\ref{thm:near-unanimity},
Theorem~\ref{thm:maltsev}, and the known fact that 
any constraint language with a constant polymorphism 
has a set function~\cite{dp99}
along with Theorem~\ref{thm:set-function}.
Otherwise, denote $D = \{ 0, 1 \}$;
by Post's lattice~\cite{bcrv03-post} it is known that 
$\pol(\Gamma)$ is contained in the clone of operations $F_u$
generated by the unary operation
mapping $0$ to $1$, and $1$ to $0$.
Let $\nae$ denote the ``not-all-equal'' relation
$\{ 0, 1 \}^3 \setminus \{ (0, 0, 0), (1, 1, 1) \}$.
It is known that $F_u$ is exactly the 
set of polymorphisms of $\nae$~\cite{bcrv04-csp},
and so by Proposition~\ref{prop:pol-complexity}
with $\Gamma_1 = \{ \nae \}$ and $\Gamma_2 = \Gamma$,
it suffices to show that $\qcspe(\{ \nae \})$ has maximal complexity.
Since $\qcsp(\{ \nae \} \cup \{ (0) \} \cup \{ (1) \})$ has
maximal complexity by~\cite{dalmau97,cks01,hemaspaandra04},
by Proposition~\ref{prop:standard-to-our}, it suffices to 
give a uniform reduction from
$\qcspe(\{ \nae \} \cup \{ (0) \} \cup \{ (1) \})$ 
to
$\qcspe(\{ \nae \})$.
Given an instance $\phi$ of
$\qcspe(\{ \nae \} \cup \{ (0) \} \cup \{ (1) \})$, we create
an instance of $\qcspe(\{ \nae \})$ as follows.
First, create two new existentially quantified variables
$c$ and $c'$, and add the extended constraint $\nae(c, c, c')$.
This extended constraint guarantees that $c$ and $c'$ have different values.
Replace each extended constraint of the form
$$(y_1 = d_1) \wedge \ldots \wedge (y_m = d_m) \Rightarrow \{ (0) \}(x)$$
with
$$(y_1 = d_1) \wedge \ldots \wedge (y_m = d_m) \Rightarrow \nae(x, x, c),$$
and, 
replace each extended constraint of the form
$$(y_1 = d_1) \wedge \ldots \wedge (y_m = d_m) \Rightarrow \{ (1) \}(x)$$
with
$$(y_1 = d_1) \wedge \ldots \wedge (y_m = d_m) \Rightarrow \nae(x, x, c').$$
A conjunction of $\nae$ constraints is satisfied by an assignment
if and only if it is also satisfied by the ``flipped'' assignment
where $0$ and $1$ are swapped.  Hence, if there is any winning strategy
for the resulting quantified formula, there is a winning strategy
where $c$ is set to $1$ and $c'$ is set to $0$.
It is easily verified
that if the variables $c, c'$ are removed from such a winning strategy,
one obtains a winning strategy for the original formula;
likewise, a winning strategy for the original formula
augmented to set $c$ to $1$ and $c'$ to $0$ yields a winning strategy
for the new formula.
\end{pf}

\section{Set functions}
\label{sect:set-functions}

This section investigates the complexity of idempotent set functions 
in the standard model, by using our existentially restricted model.
A set function $f: \wp(D) \rightarrow D$ is \emph{idempotent}
if $f(\{ d \}) = d$ for all $d \in D$.
What is known about idempotent set functions in the standard model?
Previous work~\cite{chen05-maximal} has demonstrated that
idempotent set functions $f$ give rise to two modes
of behavior in the QCSP: either $\qcsp(f)$ is $\conp$-hard,
or $\qcsp(f)$ is in $\p$.  In particular, the following is known.

\begin{definition}
Let $f: \wp(D) \rightarrow D$ be an idempotent set function.
Say that $C \subseteq D$ is coherent (with respect to $f$) if 
it is non-empty and for all non-empty $A \subseteq D$,
it holds that $f(A) \in C$ implies $A \subseteq C$.
We say that $f$ is \emph{hard} if it has two disjoint coherent sets;
otherwise, we say that $f$ is \emph{easy}.
\end{definition}

\begin{theorem} \cite{chen05-maximal}
\label{thm:set-function-classification}
Let $f: \wp(D) \rightarrow D$ be an idempotent set function.
If $f$ is hard, then $\tqcsp(f)$ is $\conp$-hard for all $t \geq 2$.
If $f$ is easy, then $\qcsp(f)$ is in $\p$.
\end{theorem}

As the complexity of easy set functions is known, our focus here is on
 the hard set functions.  Our first observation is that,
under bounded alternation, hard set functions are $\conp$-complete.

\begin{theorem}
\label{thm:set-function-standard}
If $f$ is a hard set function, then $\tqcsp(f)$ is $\conp$-complete
(for all $t \geq 2$).
\end{theorem}

\begin{pf}
Hardness for $\conp$ is immediate from Theorem~\ref{thm:set-function-classification}.
Inclusion in $\conp$ follows 
 from Theorem~\ref{thm:set-function} and
Proposition~\ref{prop:standard-to-our}.
\end{pf}

We have that for a hard set function $f$, the problem 
$\tqcsp(f)$ is in $\conp$, that is, $\qcsp(f)$ is in $\conp$ under
bounded alternation. 
This result naturally prompts the
 question of whether or not $\qcsp(f)$
is in $\conp$ under unbounded alternation.  
We are able to answer this question in the negative:
we demonstrate that such a $\qcsp(f)$ is $\pip_2$-hard,
implying that if $\qcsp(f)$ were
in $\conp$, we would have $\pip_2 = \conp$ and that the polynomial
hierarchy collapses.
We in fact show $\pip_2$-hardness of such $\qcsp(f)$ by first showing that 
extended quantified Horn formulas reduce to $\qcsp(f)$,
and then that extended quantified Horn formulas are $\pip_2$-hard.

\begin{theorem}
If $f$ is a hard set function, then $\qcsp(f)$ is $\pip_2$-hard.
\end{theorem}

\begin{pf}
Immediate from Theorems~\ref{thm:horn-to-set} and~\ref{thm:horn-pi2-hard},
proved below.
\end{pf}

\begin{theorem}
\label{thm:horn-to-set}
Let $f$ be a hard set function.
The problem of deciding the truth of extended quantified Horn formulas 
uniformly reduces to $\qcsp(f)$.
\end{theorem}

\begin{pf}
Let $C_0$ and $C_1$ be disjoint coherent sets with respect to $f$,
let $C$ be a coherent set with respect to $f$ that is not equal to $D$,
and let $c_t \in C$ be a fixed element of $C$.

Given an extended quantified Horn formula $\phi$, we create
an instance $\phi'$ of $\qcsp(f)$ as follows.
First, note that by the introduction of extra existentially quantified
variables, we can transform $\phi$ in polynomial time into another
extended quantified Horn formula in which each clause has a constant
number of literals.  We thus assume that each clause of $\phi$
has a constant number of literals.

We create a constraint invariant under $f$ for each extended Horn
clause of $\phi$.  In particular, for each extended Horn clause
$l_1 \vee \ldots \vee l_k$ of $\phi$, we create in $\phi'$ the constraint
$m_1 \vee \ldots \vee m_k$, where
\begin{itemize}
\item $m_i = (y \notin C_0)$ if $l_i = y$ and $y$ is a universally quantified variable,
\item $m_i = (y \notin C_1)$ if $l_i = \overline{y}$ and $y$ is a universally quantified variable,
\item $m_i = (x = c_t)$ if $l_i = x$ and $x$ is an existentially quantified variable, and 
\item $m_i = (x \notin C)$ if $l_i = \overline{x}$ and $x$ is an existentially quantified variable.
\end{itemize}
Let us verify that any such constraint
$M = m_1 \vee \ldots \vee m_k$ is invariant under $f$.
Let $a_1, \ldots,  a_n$ 
be assignments to the variables $V$ of $M$
satisfying $M$.  We want to show that the assignment
$a$ defined by $a(v) = f(\{ a_1(v), \ldots, a_n(v) \})$ 
for all variables $v \in V$,
also satisfies $M$.
If any $a_i$ satisfies $M$ by satisfying
an $m_j$ of the form $(y \notin C_0)$, $(y \notin C_1)$, or $(x \notin C)$,
then $a$ also satisfies the $m_j$ by the coherence of $C_0$, $C_1$, and $C$.
(Suppose for instance $a_i(y) \notin C_0$.  Then 
we have $a(y) = a(S)$ for a set $S$ including $a_i(y)$, which is not in $C_0$, 
and so by the coherence of $C_0$, we have $a(y) \notin C_0$.)
Otherwise, there exists an $m_j$ with $m_j = (x = c_t)$.  But since
the original clause $l_1 \vee \ldots \vee l_k$ was an extended
Horn clause, it contained at most one existentially quantified variable
appearing positively, and there is a unique such $m_j = (x = c_t)$.
We have $a_i(x) = c_t$ for all $i = 1, \ldots, n$ and thus, by the
idempotence of $f$, $a(x) = f( \{ c_t \}) = c_t$.

We have shown that $\phi'$ is indeed an instance of $\qcsp(f)$.
It remains to show that $\phi'$ is true if and only if $\phi$ is true.
Notice that in $\phi'$, a strategy is winning as long as it is winning
with respect to all assignments $\tau: Y \rightarrow D$ 
to the universally quantified
variables $Y$ where $\tau(y) \in C_0 \cup C_1$ for all $y \in Y$;
the sets $C_0$ and $C_1$ in $\phi'$ encode the values $0$ and $1$ 
for the universally quantified variables in $\phi$.
Let $\{ \sigma_x \}$ be a winning strategy for $\phi$.
Define $\sigma'_x$ to be equal to $c_t$ whenever $\sigma_x$ is equal to $1$,
and to be an element of $D \setminus C$ whenever $\sigma_x$ is equal to $0$.
Associating the sets $C_0$ and $C_1$ with $0$ and $1$ as discussed,
it is straightforward to verify that $\{ \sigma'_x \}$ 
is a winning strategy for $\phi'$.
Likewise, if $\{ \sigma'_x \}$ is a winning strategy for $\phi'$,
define $\sigma_x$ to be $1$ whenever $\sigma'_x$ is equal to $c_t$,
and $0$ otherwise; using the same association, 
it is straightforward to verify that
$\{ \sigma'_x \}$ is a winning strategy for $\phi$.
\end{pf}

\begin{theorem}
\label{thm:horn-pi2-hard}
The problem of deciding the truth of extended quantified Horn formulas 
(without
any bound on the number of alternations) is $\pip_2$-hard.
\end{theorem}

The proof of this theorem is inspired by the proof
of~\cite[Theorem 3.2]{bkf95}.

\begin{pf}
We first prove that the problem is $\np$-hard, then indicate how
the proof can be generalized to yield $\pip_2$-hardness.
We reduce from CNF satisfiability.
Let $\phi$ be a CNF formula over variable set $\{ y_1, \ldots, y_n \}$.
We create an extended quantified Horn formula $\phi'$
based on $\phi$ as follows.
The quantifier prefix of $\phi'$ is
$$(\exists x_1^0 \exists x_1^1 \forall y_1) \ldots (\exists x_n^0 x_n^1 \forall y_n) \exists d.$$
The clauses of $\phi'$ are as follows.
We call the following the \emph{core clauses} of $\phi'$; they do not
depend on $\phi$.
$$x_1^0 \wedge x_1^1 \Rightarrow \mathsf{False}$$
$$y_i \wedge x_{i+1}^0 \wedge x_{i+1}^1 \Rightarrow x_i^1 \mbox{ for all } 
i \in \{ 1, \ldots, n-1 \}$$
$$\overline{y_i} \wedge x_{i+1}^0 \wedge x_{i+1}^1 \Rightarrow x_i^0 \mbox{ for all } 
i \in \{ 1, \ldots, n-1 \}$$
$$y_n \wedge d \Rightarrow x_n^1$$
$$\overline{y_n} \wedge d \Rightarrow x_n^0$$

In addition, for each clause $l_1 \vee l_2 \vee l_3$ of $\phi$,
there is a clause in $\phi'$ of the form
$$\overline{l_1} \wedge \overline{l_2} \wedge \overline{l_3} \Rightarrow d$$

We claim that $\phi$ is satisfiable if and only if $\phi'$ is true.

\paragraph{$\phi$ is unsatisfiable implies $\phi'$ is false.}
Suppose that $\phi$ is unsatisfiable.  
It is straightforward to verify that
the following clauses can be \emph{derived} from the clauses for $\phi'$, 
by which we mean that
 any winning strategy for $\phi'$ must 
satisfy the following clauses.

We can derive
$$z_1 \wedge \ldots \wedge z_n \Rightarrow d$$
for any choice of $z_1, \ldots, z_n$ with $z_i \in \{ y_i, \overline{y_i} \}$
because any assignment to the variables $\{ y_1, \ldots, y_n \}$
falsifies a clause, and hence makes the left-hand side of a clause
$$\overline{l_1} \wedge \overline{l_2} \wedge \overline{l_3} \Rightarrow d$$
true.

Now, by induction (starting from $n-1$), it can be shown that for all
$k = n-1, \ldots, 0$, the clauses
$$z_1 \wedge \ldots \wedge z_k \Rightarrow x_{k+1}^0$$
$$z_1 \wedge \ldots \wedge z_k \Rightarrow x_{k+1}^1$$
can be derived,
for any choice of $z_1, \ldots, z_k$ with $z_i \in \{ y_i, \overline{y_i} \}$.
This is done by using the core clauses (other than the first core clause).

By this induction,
we obtain that $x_1^0$ and $x_1^1$ can be derived (by setting $k = 0$);
using the clause
$$x_1^0 \wedge x_1^1 \Rightarrow \mathsf{False}$$
we can then derive $\mathsf{False}$.

\paragraph{$\phi$ is satisfiable implies $\phi'$ is true.}
Suppose that $\phi$ is satisfiable via the assignment
$f: \{ y_1, \ldots, y_n \} \rightarrow \{ 0, 1 \}$. 
We describe a winning strategy for $\phi'$.  (Note that we use the notation
$\overline{0} = 1$ and $\overline{1} = 0$.)
The strategy sets $x_i^{\overline{f(y_i)}}$ to be true,
and $x_i^{f(y_i)}$ to be false 
if and only if for every
$j < i$, the variable $y_j$ has been set to $f(y_j)$.
Also, the strategy sets $d$ to be false
if and only if every variable $y_j$ has been set to $f(y_j)$.

Consider an arbitrary assignment 
$g: \{ y_1, \ldots, y_n \} \rightarrow \{ 0, 1 \}$.  We wish to show
that in the formula $\phi'$, when the variables are set according to 
$g$ and the described strategy, all clauses are indeed satisfied.

The first core clause
$$x_1^0 \wedge x_1^1 \Rightarrow \mathsf{False}$$
is satisfied since $x_0^{\overline{f(y_0)}}$ is set to be false.

Now consider a core clause of the form
$$y_i \wedge x_{i+1}^0 \wedge x_{i+1}^1 \Rightarrow x_i^1$$
with $i \in \{ 1, \ldots, n-1 \}$.
Assume that
$x_i^1$ is false and $y_i$ is true under $g$
(if not, the clause is clearly satisfied).
The falsity of $x_i^1$ implies, by the definition of our strategy,
that $f(y_i) = 1$ as well as that for all $j < i$,
the variable $y_j$ has been set according to $f$, that is, $f(y_j) = g(y_j)$.
Now, since $f(y_i) = 1 = g(y_i)$, it indeed holds that
for all $j \leq i$, 
the variable $y_j$ has been set according to $f$, that is, $f(y_j) = g(y_j)$.
By the definition of our
strategy, then, $x_i^{f(y_i)}$ is set to false, and the clause is satisfied.
Similar reasoning applies to the remaining three types of core clauses.

Now consider the clauses of the form
$$\overline{l_1} \wedge \overline{l_2} \wedge \overline{l_3} \Rightarrow d$$
If $d$ is set to false, then $g$ must be equal to $f$, which satisfies $\phi$.
Thus,
$$\overline{l_1} \wedge \overline{l_2} \wedge \overline{l_3}$$
which is the negation of a clause from $\phi$, is false under $g$.

\paragraph{$\pip_2$-hardness.}
We now indicate how to extend the given proof to show that 
extended quantified Horn formulas are in fact $\pip_2$-hard.
Let $\phi = \forall w_1 \ldots \forall w_m \exists y_1 \ldots \exists y_n \cons$
be a quantified boolean formula where $\cons$ is a 3-CNF.
The extended quantified Horn formula $\phi'$ 
that we create has quantifier prefix
$$(\forall w_1 \ldots \forall w_m)(\exists x_1^0 \exists x_1^1 \forall y_1) \ldots (\exists x_n^0 x_n^1 \forall y_n) \exists d.$$
The clauses of $\phi'$ are defined as above.  
The key point is that, under any assignment to the variables
$\{ w_1, \ldots, w_m \}$, the formula $\phi$ is true if and only if
the formula $\phi'$ is true, by using the reasoning in the
$\np$-hardness proof.
\end{pf}

Theorem~\ref{thm:horn-pi2-hard} has an interesting implication
for our model of existentially restricted quantified constraint satisfaction.
This paper has focused mainly on proving that for certain constraint
languages $\Gamma$, it holds that $\tqcspe(\Gamma)$ is in $\conp$,
that is, 
the \emph{bounded alternation formulas} for $\qcspe(\Gamma)$ are in
$\conp$.  This theorem implies that such results, in general,
can \emph{not} be extended to the case of unbounded alternation.
In particular, we have such a constraint language $\Gamma$
whose $\qcspe(\Gamma)$ complexity--in the case of unbounded alternation--is 
\emph{not} in $\conp$, unless 
$\conp = \pip_2$ and
the polynomial hierarchy collapses.

\begin{theorem}
\label{thm:extended-quantified-horn-complexity}
Let $\Gamma_H$ be the constraint language defined in 
Example~\ref{ex:extended-quantified-horn}.
It holds that $\tqcspe(\Gamma_H)$ is in $\conp$
for all $t \geq 2$,
but $\qcspe(\Gamma_H)$ is $\pip_2$-hard.
\end{theorem}


\begin{pf}
Inclusion of $\tqcspe(\Gamma_H)$ in $\conp$ is discussed in 
Example~\ref{ex:extended-quantified-horn}; there,
it is also pointed out that $\qcspe(\Gamma_H)$ is equivalent to the
problem of deciding the truth of extended quantified Horn formulas,
so the $\pip_2$-hardness of $\qcspe(\Gamma_H)$ follows from 
Theorem~\ref{thm:horn-pi2-hard}.
\end{pf}

\section{Discussion}

\subsection{Comparison with the standard model}

As we have discussed in the introduction,
 our new model 
and the standard model are both natural generalizations of the CSP model.
However, we would like to argue here that our new model is,
in certain respects, 
 a more faithful generalization
of the CSP model.

In the CSP model, it is possible to use algebraic notions
such as subalgebra and homomorphic image to study the complexity of
constraint languages \cite{bkj00}.  We have shown
that these notions can also be used to study complexity
in our new model (Theorem~\ref{thm:sub-alg-hom}), indicating
that our model is algebraically robust.  
In contrast, the first property
of Theorem~\ref{thm:sub-alg-hom} does \emph{not} hold in the standard
model: 
using the results of \cite[Chapter 5]{chen05-thesis}, 
it is easy to construct
a semilattice $\alga$ having a subalgebra $\algb$ such that
$\qcsp(\alga)$ is polynomial-time decidable, but $\qcsp(\algb)$ is 
$\conp$-hard.


Along these lines, it is known that there are constraint languages with
trivial CSP complexity--where all CSP instances are satisfiable
by a mapping taking all variables to the same value--but 
maximal QCSP  complexity,
under the standard model.\footnote{
An example of a constraint language having
the mentioned properties is the constraint language 
over the domain $\{ 0, 1 \}$
consisting of all arity four relations
that contain the all-0 tuple $(0,0,0,0)$.
This is PSPACE-complete in the alternation-unbounded standard model
by the results~\cite{dalmau97,cks01}, 
and complete for the various levels of the
polynomial hierarchy in the alternation-bounded standard model
by the result~\cite{hemaspaandra04}.
}
This ``wide complexity gap'' phenomenon of the standard model
does not appear to occur in our new model.
To understand why, it is didactic to consider the case of 
QCSP instances with quantifier type $\forall \exists$.
Fix any constraint language giving rise to a 
polynomial-time tractable case of the CSP.  In our new model,
$\forall \exists$ instances over the constraint language are
immediately seen to be in coNP.
The argument is simple: 
 once the universal 
variables are instantiated, the result is an instance of the CSP
over the constraint language,
which can be decided in polynomial time.
In contrast, in the standard model, 
$\forall \exists$ instances
over the constraint language may be $\pip_2$-complete,
that is, maximally hard given the quantifier prefix.
The argument given for our new model does not apply:
even if the constraints of a $\forall \exists$ instance 
are originally over the constraint language, after
the universal variables have been instantiated with values,
new constraints may be created that are not over the original
constraint language.  The key point is that while instantiation of universally
quantified variables may ``disrupt'' the constraint language in the
standard model, it does not do so in our new model.  
All in all, the faithfulness of our model to the original CSP model
affords
a fresh opportunity to 
enlarge the repertoire of positive QCSP complexity results,
by way of extending existing CSP tractability results.

\subsection{Conclusions}

We have introduced and studied a new model for restricting the QCSP,
a generally intractable problem.  
We presented powerful technology for proving coNP-inclusion results
in this new model, under bounded alternation, and have applied this
technology to a variety of constraint languages.
We also derived new results on the standard model using
results on our new model, in particular, new results on the complexity
of constraint languages having a set function polymorphism.
In addition, we demonstrated that, in general, coNP-inclusion results
for our new model in the case of bounded alternation, 
can \emph{not} be extended to the case of unbounded alternation.

One interesting direction for future research is to classify the
complexity of all constraint languages in our new model under
bounded alternation.  At this moment, a plausible conjecture is that
all constraint languages that are tractable in the CSP model are
in coNP in our new model, under bounded alternation.
A second direction is to investigate further the case of unbounded
alternation.  In particular, one could investigate the 
unbounded-alternation complexity
of constraint languages that are known to be in coNP under bounded
alternation; Theorem~\ref{thm:extended-quantified-horn-complexity}
represents one step in this direction.

\paragraph{Acknowledgements.}  The author wishes to thank
Manuel Bodirsky, V\'{\i}ctor Dalmau, Riccardo Pucella, and Ryan Williams
for helpful comments.


\appendix

\section{Proof of Theorem \ref{thm:polysize-proofs}}

Our first step is to show completeness of the proof system,
that is, if
$\phi$ is false, then
$\phi, \top \vdash \bot$.
We accomplish this in a sequence of lemmas.
Throughout, we assume that the quantified formulas under discussion
are all instances of $\qcspe(\Gamma)$, and we fix a fingerprint scheme
for $\Gamma$.

\begin{lemma}
\label{lemma:derive}
For any quantified formula $\phi$ with $t$ existential blocks
and any fingerprint application $F\lf X \rf$ for $\phi$ with
$X \subseteq X_1^{\phi}$, there exists a proof of size $t$ that
$\phi, F\lf X \rf \vdash F'\lf X_1^{\phi} \rf$ for some fingerprint 
$F'$ with $\pi_{|X|} F' \sqsubseteq F$.
\end{lemma}

\begin{pf}
The proof is by induction on $t$.  If $t = 1$, the proof consists
of one instance of the first proof rule; the
$\pi_{|X|} \rel(F') \sqsubseteq F$ 
criterion
(of the lemma)
 holds because of the
``progress'' requirement on inference algorithms.
If $t > 1$, then let $g: Y_1^{\phi} \rightarrow D$ be any mapping.
By induction, there is a proof of size $t-1$ that
$\phi[g], F\lf X \rf \vdash F'\lf X_1^{\phi} \cup X_2^{\phi} \rf$ 
for some fingerprint 
$F''$ with 
$\pi_{|X|} F'' \sqsubseteq F$.
Applying the third proof rule, we obtain a proof of size $t$ that
$\phi, F\lf X \rf \vdash  (\pi_{|X_1^{\phi}|} F'')\lf X_1^{\phi} \rf$.
Set $F' = (\pi_{|X_1^{\phi}|} F'')$.
By Definition~\ref{fingerprint-collection}, 
we have 
$\pi_{|X|} F' = \pi_{|X|} F''$.
Since $\pi_{|X|} F'' \sqsubseteq F$, we have the lemma.
\end{pf}

We say that a fingerprint application $F\lf X \rf$ is \emph{stable} for
a quatified formula $\phi$ if $X = X_1^{\phi}$, 
$F \neq \bot$, it holds that
$\phi, \top \vdash F\lf X \rf$, and
$F' \sqsubseteq F$ implies
$\phi, \top \not\vdash F'\lf X \rf$.
That is, a fingerprint application is stable for $\phi$ if it can be derived,
but no fingerprint application ``lower'' than it can be derived.

\begin{lemma}
\label{lemma:has-stable}
If $\phi$ is a quantified formula such that
$\phi, \top \not\vdash \bot$, then $\phi$ has a stable fingerprint application.
\end{lemma}

\begin{pf}
Immediate from the definition of stable fingerprint application and Lemma
\ref{lemma:derive}, which implies that for \emph{some} fingerprint application
$F\lf X \rf$ with $X = X_1^{\phi}$,
it holds that $\phi, \top \vdash F\lf X \rf$. 
\end{pf}

Let us say that a fingerprint $F$
\emph{extends} another fingerprint $F'$
if $\arity(F) \geq \arity(F')$ and 
$\pi_{\arity(F')} F = F'$.

\begin{lemma}
\label{lemma:fingerprint-extends}
If $\phi$ is a quantified formula with stable fingerprint application $F\lf X\rf$
and having two or more existential blocks,
then for any mapping $g: Y_1^{\phi} \rightarrow D$, the quantified
formula $\phi[g]$ has a stable fingerprint application
$F'\lf X'\rf$ where $F'$  extends $F$.
\end{lemma}

\begin{pf}
By assumption, 
we have $\phi, \top \vdash F\lf X \rf$, with $X = X_1^{\phi}$.
Let $g: Y_1^{\phi} \rightarrow D$ be any mapping.
By Lemma~\ref{lemma:derive}, there exists a fingerprint
$F_2$ of arity $|X_1^{\phi} \cup X_2^{\phi}|$ having the properties that
$\phi[g], F\lf X \rf \vdash F_2\lf X' \rf$ and
$\pi_{|X|} F_2 \sqsubseteq F$, where
$X' = |X_1^{\phi} \cup X_2^{\phi}|$.
Now let $F'$ be a ``minimal'' fingerprint with the above properties,
that is, a fingerprint
such that for no other $F_2$ having the above properties does it hold
that $F_2 \sqsubseteq F'$.

We claim that $F'\lf X' \rf$ is a stable fingerprint application for
$\phi[g]$ where $F'$ extends $F$.
It suffices to show that $F'$ extends $F$.
Suppose not; then $\pi_{|X|} F' \sqsubseteq F$
and $\pi_{|X|} F' \neq F$.
By the third rule of the proof system,
$\phi, F\lf X \rf \vdash (\pi_{|X|} F')\lf X \rf$
and by the second rule, $\phi, \top \vdash (\pi_{|X|} F')\lf X \rf$,
contradicting that $F\lf X \rf$ is stable for $\phi$.
\end{pf}

\begin{lemma}
\label{lemma:stable-true}
If $\phi$ is a quantified formula with a stable fingerprint application,
then $\phi$
is true.
\end{lemma}

\begin{pf}
We prove the following claim: if $F\lf X_1^{\phi} \rf$
is a stable fingerprint application for $\phi$, then $\phi$ is true
via the assignment 
$f: X_1^{\phi} \rightarrow D$ defined 
by the construction mapping $\algcons$ and the fingerprint application
$F\lf X_1^{\phi} \rf$, as in Definition~\ref{def:fingerprint-scheme}.

We prove this claim by induction on the number $t$ of existential
blocks in $\phi$.  Let $\cons$ denote the constraints of $\phi$.

When $t = 1$, it is straightforward to verify that any 
 fingerprint application derivable from $\phi, \top$, and hence any
stable
fingerprint application, is suitable
for $\cons$.  Note that only the first two proof rules can be used
to perform derivations from $\phi, \top$.

When $t > 1$, by Lemma~\ref{lemma:fingerprint-extends}
 we have that for all $g: Y_1^{\phi} \rightarrow D$, there is a 
stable fingerprint application $F_g \lf X_1^{\phi[g]} \rf$ for $\phi[g]$
where
$F_g$ extends $F$.
By induction, for all $g$, the quantified formula
$\phi[g]$ is true via the assignment
$f_g: X_1^{\phi[g]} \rightarrow D$ defined 
by $\algcons$ and 
$F_g \lf X_1^{\phi[g]} \rf$, as in Definition~\ref{def:fingerprint-scheme}.
The claim follows from the observation that 
for all $g: Y_1^{\phi} \rightarrow D$, 
the restriction of $f_g$ to $X_1^{\phi}$ is equal to 
$f: X_1^{\phi} \rightarrow D$.
\end{pf}

We are now able to observe that the proof system is complete.

\begin{lemma}
\label{lemma:complete}
If the quantified formula $\phi$ is false, then $\phi$ has a proof of falsity
(that is, $\phi, \top \vdash \bot$).
\end{lemma}

\begin{pf}
If $\phi, \top \not\vdash \bot$, then by 
Lemma~\ref{lemma:has-stable}, the formula $\phi$ has a 
stable fingerprint application; it follows from
Lemma~\ref{lemma:stable-true} that the formula $\phi$ is true.
\end{pf}

However, we want to show something stronger than Lemma~\ref{lemma:complete}:
that every false quantified formula has a succinct proof of falsity.
The following lemma is key; roughly speaking, it shows that for 
alternation-bounded formulas, if there is a proof of falsity at all,
then there is a succinct proof of falsity.

\begin{lemma}
\label{lemma:succinct}
For each $t \geq 1$, there exists a polynomial $p_t$ such that if
$\phi$ has $t$ existential blocks, $n$ existential variables, and
$\phi, F\lf X \rf \vdash F'\lf X_1^{\phi} \rf$, then there is a proof of
$\phi, F\lf X \rf \vdash F'\lf X_1^{\phi} \rf$ of size bounded above by 
$p_t(n)$.
\end{lemma}

\begin{pf}
We first observe that the proof system has the following monotonicity property:
if $\phi, F \lf X \rf \vdash F' \lf X' \rf$, then 
$\pr_{|X|} \rel(F') \subseteq \rel(F)$.  This is straightforward to verify.

We prove the lemma by induction on $t$.  When $t = 1$, 
inspecting
a proof of $\phi, F\lf X \rf \vdash F'\lf X_1^{\phi} \rf$,
it can be seen that
only the first two proof rules are applicable, and 
by monotonicity
there must be distinct
fingerprints $F_1, \ldots, F_m$ with $F_m = F'$ and
$F_1 \sqsubseteq \cdots \sqsubseteq F_m$
such that

$$\Rule{}{\phi, F\lf X \rf \vdash F_1\lf X_1^{\phi} \rf}$$

and

$$\Rule{}{\phi, F_k\lf X_1^{\phi} \rf \vdash F_{k+1}\lf X_1^{\phi} \rf}$$

for all $k = 1, \ldots, m-1$.
The above proofs can be combined into a proof of
$\phi, F\lf X \rf \vdash F'\lf X_1^{\phi} \rf$ using no more than
$(m-1)$ applications of the second rule.  We thus obtain a proof of size
$m + (m-1)$, which is polynomial in $n = |X_1^{\phi}|$ because of
the requirement that in a fingerprint collection, chains are of
polynomial length.

When $t > 1$, 
inspecting a proof of $\phi, F\lf X \rf \vdash F'\lf X_1^{\phi} \rf$
and using monotonicity, by induction on the structure of the proof
it can be seen that there must be 
mappings $g_1, \ldots, g_m: Y_1^{\phi} \rightarrow D$,
distinct
fingerprints $F_1, \ldots, F_m$ of arity $|X_1^{\phi}|$ with $F_m = F'$,
and
fingerprints $F'_1, \ldots, F'_m$,
 where
$$F_1 \sqsubseteq \cdots \sqsubseteq F_m$$
and such that

$$\Rule{\phi[g_1], F \lf X \rf  \vdash  F'_1 \lf X_1^{\phi[g_1]} \rf  \qquad  
                                          X_1^{\phi[g_1]} \supseteq X_1^{\phi}}
       {\phi,      F \lf X \rf  \vdash 
                    F_1\lf X_1^{\phi} \rf}$$

and

$$\Rule{\phi[g_k], F_k \lf X_1^{\phi} \rf  \vdash  F'_{k+1} \lf X_1^{\phi[g_k]} \rf  \qquad  
                                          X_1^{\phi[g_k]} \supseteq X_1^{\phi}}
       {\phi,      F_k \lf X_1^{\phi} \rf  \vdash 
                    F_{k+1}\lf X_1^{\phi} \rf}$$

for all $k = 1, \ldots, m-1$.  As before, 
the above proofs can be combined into a proof of
$\phi, F\lf X \rf \vdash F'\lf X_1^{\phi} \rf$ using no more than
$(m-1)$ applications of the second rule.  By induction, 
the hypothesis of each rule instance given above has a proof of size
$p_{t-1}(n)$.
We thus obtain a proof of size bounded above by
$m(p_{t-1}(n) + 1) + (m-1)$.
This expression is polynomial in $n$:  $m$ is
polynomial in $|X_1^{\phi}| \leq n$ because of 
the requirement that in a fingerprint collection, chains are of
polynomial length.
\end{pf}

\begin{pf} (Theorem \ref{thm:polysize-proofs})
By Lemma~\ref{lemma:complete},
for all false formulas $\phi$ of $\tqcspe(\Gamma)$,
it holds that $\phi, \top \vdash \bot$.
By Lemma~\ref{lemma:succinct}, there is a proof of
$\phi, \top \vdash \bot$ of size bounded above by $p_t(n)$ where $n$ is the
number of variables of $\phi$, and $p_t$ is a polynomial.
\end{pf}

\section{Proof of Theorem \ref{thm:near-unanimity}}

\begin{pf}
Let $\Gamma$ be a constraint language over domain $D$
having $f: D^k \rightarrow D$ as near-unanimity polymorphism.
By Theorem \ref{thm:in-conp}, it suffices to show that
$\Gamma$ has a fingerprint scheme.

We first describe the fingerprint collection.

The fingerprints of arity $n$ are sets of constraints
on $\{ v_1, \ldots, v_n \}$.  In particular, a fingerprint of arity $n$
contains exactly one constraint for each possible variable set
of size less than or equal to $k$; formally,
there is a constraint over each variable tuple of the form
$(v_{i_1}, \ldots, v_{i_m})$
with $1 \leq i_1 < \cdots < i_m \leq n$ and $1 \leq m \leq k$.

The relation specified by a fingerprint $F$ is the set of all tuples
$(d_1, \ldots, d_n)$ such that the mapping $v_i \rightarrow d_i$
satisfies all constraints in $F$.

The projection function $\pi$, given a fingerprint $F$ and $k$,
 simply projects all constraints onto
the variables $\{ v_1, \ldots, v_k \}$.

For two fingerprints $F, F'$ of arity $n$, we define
$F \sqsubseteq F'$ if and only if 
for each constraint
$R(v_{i_1}, \ldots, v_{i_m})$ in $F$, the corresponding constraint
$R'(v_{i_1}, \ldots, v_{i_m})$ in $F'$ satisfies $R \subseteq R'$.

Chains are of polynomial length, since the length of a chain is bounded
above by the total number of tuples that can appear in a fingerprint.
This total number
is equal to the number of constraints,
${n \choose k} + {n \choose k-1} + \cdots + {n \choose 1}$,
 times $|D|^k$, which upper bounds the number of tuples in each constraint;
this is clearly polynomial in $n$, for fixed $D$ and $k$.

We now describe the fingerprint scheme.

The inference algorithm, given a fingerprint application
$F\lf v_1, \ldots, v_k \rf$ and a conjunction of constraints $\cons$
over $\Gamma$ and $\{ v_1, \ldots, v_n \}$, establishes
strong $k$-consistency on the constraints $\cons \cup F$
to obtain $\cons'$; 
the result is the fingerprint $F'$ of arity $n$
where each constraint 
$R(v_{i_1}, \ldots, v_{i_m})$ 
contains the solutions to $\cons'$
restricted to $\{ v_{i_1}, \ldots, v_{i_m} \}$.
Please see \cite{jcc98} for the definition of strong $k$-consistency
and solution.

We define the construction mapping as follows.  Let
$F$ be a fingerprint of arity $n$, with $\rel(F) \neq \emptyset$, 
that is suitable for a conjunction
of constraints $\cons$.  Due to the definition of our inference algorithm,
we know that $F$ is strongly $k$-consistent, and that any 
assignment satisfying the constraints in $F$ also satisfies $\cons$.
The construction mapping is defined inductively: it simply computes
the mapping $a: \{ v_1, \ldots, v_{n-1} \} \rightarrow D$ defined
by $a(v_i) = \algcons(\pi_i F)$, and then outputs a value $d$
such that the extension of $a$ mapping $v_n$ to $d$ satisfies the
constraints in $F$; such a value is guaranteed to exist by the 
fact that $F$ is strongly $k$-consistent and
\cite[Theorem 3.5]{jcc98}.  Notice that  if
$F\lf v_1, \ldots, v_n \rf$ is a fingerprint application suitable for
$\cons$, then the mapping defined in Definition~\ref{def:fingerprint-scheme}
satisfies
the constraints in $F$, which in turn (as we pointed out) implies
that this mapping satisfies the constraints of $\cons$, as desired.
\end{pf}

\section{Proof of Theorem \ref{thm:maltsev}}

\newcommand{\lgen}{\langle}
\newcommand{\rgen}{\rangle_{\phi}}
\newcommand{\sig}{\mathsf{Sig}}

\begin{pf}
Let $\Gamma$ be a constraint language over domain $D$
having $\phi: D^3 \rightarrow D$ as Mal'tsev polymorphism.
By Theorem \ref{thm:in-conp}, it suffices to show that
$\Gamma$ has a fingerprint scheme.

We assume basic familiarity with the paper~\cite{dalmau04-maltsev},
and use the terminology of that paper.  

We first describe the fingerprint collection.

The fingerprints are the relations 
that are compact representations, that is, the relations
$F \subseteq D^k$ having the property
that $F$ is a compact representation of some relation.

The relation specified by a fingerprint $F$ is 
$\rel(F) = \lgen F \rgen$
(the notation $\lgen F \rgen$ denotes the smallest relation containing
$F$ and closed under $\phi$).
Observe that any such fingerprint $F$ can be represented\footnote{
Other than in this instance,  all uses of the term ``representation''
within this proof
will refer to the notion of representation defined
in~\cite{dalmau04-maltsev}.
} 
in size
polynomial in its arity $n$, since for any compact representation $F$,
it holds that $|F| \leq 2 * n * |D|^2$, 
since $F$ has at most two tuples for each element of some signature, and
$n * |D|^2$ is an upper bound on the size of a signature.

The projection function $\pi$,
given a fingerprint $F$ and $k$, simply yields
$\pr_k F$.

For two fingerprints $F, F'$ of arity $n$, we define $F \sqsubseteq F'$
if and only if $\rel(F) \subseteq \rel(F')$.

Chains are of polynomial length: suppose
 $F, F'$ are fingerprints with 
$\rel(F) \subsetneq \rel(F')$.  
Then $\sig_{\rel(F)} \subseteq \sig_{\rel(F')}$; 
this follows immediately
from the definition of signature.  
But in fact it holds that
$\sig_{\rel(F)} \subsetneq \sig_{\rel(F')}$, for if
$\sig_{\rel(F)} = \sig_{\rel(F')}$, then
$\rel(F)$ is a representation of $\rel(F')$, and it would follow
from \cite[Lemma 1]{dalmau04-maltsev} that 
$\lgen \rel(F) \rgen = \rel(F)$ was equal to $\rel(F')$.
Since signatures of arity $n$ are subsets of a set with $n * |D|^2$
elements, chains are of polynomial length.

We now describe the fingerprint scheme.

The inference algorithm, given a fingerprint application
$F \lf x_1, \ldots, x_k \rf$ and a conjunction of constraints
$\cons$ over variables $\{ x_1, \ldots, x_n \}$, first
computes from $F$ a fingerprint $F_e$ such that
$\rel(F_e) = \rel(F) \times D^{n-k}$.
Then, starting from $F_e$, the constraints of $\cons$
are processed by {\tt Next} one by one 
(as in the algorithm {\tt Solve}) in order to 
obtain a fingerprint $F'$ 
 such that the tuples of
$\lgen F' \rgen$ are exactly the satisfying assignments of 
$\cons$ that also satisfy $F_e \lf x_1, \ldots, x_k \rf$.
Notice that $F' \sqsubseteq F_e$, and so
$\pi_k F' \sqsubseteq \pi_k F_e = F$.

The construction mapping is defined inductively.
Given a fingerprint $F$ of arity $n$ with $\rel(F) \neq \emptyset$,
it simply computes the tuple
$\overline{t} = (\algcons(\pi_1 F), \ldots, \algcons(\pi_{n-1} F))$;
by induction, we may assume that this tuple is in $\pr_{n-1} \rel(F)$.
There thus exists an element $d$ such that $(\overline{t}, d)$ is in $\rel(F)$,
which is the output of the mapping.
\end{pf}

\bibliographystyle{plain}
\bibliography{csp}

\begin{thebibliography}{10}

\bibitem{bcrv03-post}
E.~B{\"{o}}hler, N.~Creignou, S.~Reith, and H.~Vollmer.
\newblock Playing with boolean blocks, part {I}: {P}ost's lattice with
  applications to complexity theory.
\newblock {\em ACM SIGACT-Newsletter}, 34(4):38--52, 2003.

\bibitem{bcrv04-csp}
E.~B{\"{o}}hler, N.~Creignou, S.~Reith, and H.~Vollmer.
\newblock Playing with boolean blocks, part {I}{I}: constraint satisfaction
  problems.
\newblock {\em ACM SIGACT-Newsletter}, 35(1):22--35, 2004.

\bibitem{bbkj03}
F.~B\"{o}rner, A.~Bulatov, A.~Krokhin, and P.~Jeavons.
\newblock Quantified constraints: Algorithms and complexity.
\newblock In {\em Computer Science Logic 2003}, 2003.

\bibitem{bulatov02-twosemilattices}
Andrei Bulatov.
\newblock Combinatorial problems raised from 2-semilattices.
\newblock Manuscript.

\bibitem{bulatov02-dichotomy}
Andrei Bulatov.
\newblock A dichotomy theorem for constraints on a three-element set.
\newblock In {\em Proceedings of 43rd {IEEE Symposium on Foundations of
  Computer Science}}, pages 649--658, 2002.

\bibitem{bulatov02-maltsev}
Andrei Bulatov.
\newblock Malt'sev constraints are tractable.
\newblock Technical Report PRG-RR-02-05, Oxford University, 2002.

\bibitem{bulatov03}
Andrei Bulatov.
\newblock Tractable conservative constraint satisfaction problems.
\newblock In {\em Proceedings of 18th {IEEE Symposium on Logic in Computer
  Science (LICS '03)}}, pages 321--330, 2003.

\bibitem{bkj00}
Andrei Bulatov, Andrei Krokhin, and Peter Jeavons.
\newblock Constraint satisfaction problems and finite algebras.
\newblock In {\em Proceedings 27th International Colloquium on Automata,
  Languages, and Programming -- ICALP'00}, volume 1853 of {\em Lecture Notes In
  Computer Science}, pages 272--282, 2000.

\bibitem{bkj01}
Andrei Bulatov, Andrei Krokhin, and Peter Jeavons.
\newblock The complexity of maximal constraint languages.
\newblock In {\em {ACM} Symposium on Theory of Computing}, pages 667--674,
  2001.

\bibitem{bkf95}
Hans~Kleine B{\"{u}}ning, Marek Karpinski, and Andreas Fl{\"{o}}gel.
\newblock Resolution for quantified boolean formulas.
\newblock {\em Information and Computation}, 117(1):12--18, 1995.

\bibitem{chen05-thesis}
Hubie Chen.
\newblock {\em The Computational Complexity of Quantified Constraint
  Satisfaction}.
\newblock PhD thesis, Cornell University, August 2004.

\bibitem{chen04-twosemilattices}
Hubie Chen.
\newblock Quantified constraint satisfaction and 2-semilattice polymorphisms.
\newblock In {\em CP}, 2004.

\bibitem{chen05-maximal}
Hubie Chen.
\newblock Quantified constraint satisfaction, maximal constraint languages, and
  symmetric polymorphisms.
\newblock In {\em STACS}, 2005.

\bibitem{cd04-slaac}
Hubie Chen and Victor Dalmau.
\newblock ({S}mart) look-ahead arc consistency and the pursuit of {CSP}
  tractability.
\newblock In {\em Principles and Practice of Constraint Programming - CP 2004},
  Lecture Notes in Computer Science. Springer-Verlag, 2004.

\bibitem{cks01}
Nadia Creignou, Sanjeev Khanna, and Madhu Sudan.
\newblock {\em Complexity Classification of Boolean Constraint Satisfaction
  Problems}.
\newblock SIAM Monographs on Discrete Mathematics and Applications. Society for
  Industrial and Applied Mathematics, 2001.

\bibitem{dalmau97}
Victor Dalmau.
\newblock Some dichotomy theorems on constant-free quantified boolean formulas.
\newblock Technical Report LSI-97-43-R, Llenguatges i Sistemes
  Inform{\`{a}}tics - Universitat Polit{\`{e}}cnica de Catalunya, 1997.

\bibitem{dalmau04-maltsev}
Victor Dalmau.
\newblock Mal'tsev constraints made simple.
\newblock ECCC technical report, 2004.

\bibitem{dp99}
Victor Dalmau and Justin Pearson.
\newblock Closure functions and width 1 problems.
\newblock In {\em CP 1999}, pages 159--173, 1999.

\bibitem{fv98}
Tom{\'{a}}s Feder and Moshe~Y. Vardi.
\newblock The computational structure of monotone monadic snp and constraint
  satisfaction: A study through datalog and group theory.
\newblock {\em SIAM J. Comput.}, 28(1):57--104, 1998.

\bibitem{hemaspaandra04}
Edith Hemaspaandra.
\newblock Dichotomy theorems for alternation-bounded quantified boolean
  formulas.
\newblock {\em CoRR}, cs.CC/0406006, 2004.

\bibitem{jeavons98}
Peter Jeavons.
\newblock On the algebraic structure of combinatorial problems.
\newblock {\em Theoretical Computer Science}, 200:185--204, 1998.

\bibitem{jcc98}
Peter Jeavons, David Cohen, and Martin Cooper.
\newblock Constraints, consistency, and closure.
\newblock {\em Articial Intelligence}, 101(1-2):251--265, 1998.

\bibitem{jcg97}
P.G. Jeavons, D.A. Cohen, and M.~Gyssens.
\newblock Closure properties of constraints.
\newblock {\em Journal of the ACM}, 44:527--548, 1997.

\bibitem{schaefer78}
Thomas~J. Schaefer.
\newblock The complexity of satisfiability problems.
\newblock In {\em Proceedings of the ACM Symposium on Theory of Computing
  (STOC)}, pages 216--226, 1978.

\end{thebibliography}

\end{document}